\documentclass[aps, prd, amsmath, floats, floatfix, twocolumn, superscriptaddress, showpacs]{revtex4-1}
\usepackage{amsmath}    
\usepackage{amssymb}
\usepackage{amsfonts}
\usepackage{amsthm}
\usepackage{braket}
\usepackage{bm}
\usepackage{color}      
\usepackage{dcolumn}
\usepackage[dvipsnames]{xcolor}
\usepackage{epsfig}
\usepackage{graphicx}   
\usepackage{graphics}
\usepackage[latin1]{inputenc}
\usepackage{latexsym}
\usepackage{mathrsfs}
\usepackage{rotating}
\usepackage{setspace}
\usepackage{siunitx}
\usepackage{subfigure}  
\usepackage{url}
\usepackage{verbatim}	
\usepackage{hyperref}   
\usepackage{float}
\hypersetup{
	colorlinks=true,
	urlcolor=blue,
	citecolor=blue,
	linkcolor=red
}

\usepackage{xspace} 
\usepackage[toc,page]{appendix}
\usepackage{array}

\DeclareMathOperator{\Tr}{Tr}
\makeatletter
\newcommand*{\rom}[1]{\expandafter\@slowromancap\romannumeral #1@}

\newcommand{\FP}{{\mbox{\tiny FP}}}
\newcommand{\pix}{{\rm pix}}
\newcommand{\In}{{\mbox{\tiny in}}}
\newcommand{\out}{{\mbox{\tiny out}}}
\newcommand{\interact}{{\mbox{\tiny int}}}
\newcommand{\mirror}{{\mbox{\tiny M}}}

\newcommand{\AdS}{{\mbox{AdS}}}

\makeatother
\allowdisplaybreaks

\newcommand{\TAPIR}{Theoretical Astrophysics 350-17, California Institute of Technology, Pasadena, CA 91125, USA}
\newcommand{\WB}{Walter Burke Institute for Theoretical Physics, California Institute of Technology, Pasadena, CA 91125, USA}

\begin{document}
	\title{Interferometer Response to Geontropic Fluctuations}
	\author{Dongjun Li}
	\email{dlli@caltech.edu}
	\affiliation{\WB}
	\affiliation{\TAPIR}
    
    \author{Vincent S. H. Lee}
    \email{szehiml@caltech.edu}
    \affiliation{\WB}
    
    \author{Yanbei Chen}
    \email{yanbei@caltech.edu}
    \affiliation{\TAPIR}
    
    \author{Kathryn M. Zurek}
    \email{kzurek@caltech.edu}
    \affiliation{\WB}

	\date{\today}
	
\preprint{CALT-TH-2022-033}
	
\begin{abstract}
    We model vacuum fluctuations in quantum gravity with a scalar field, characterized by a high occupation number, coupled to the metric. The occupation number of the scalar is given by a thermal density matrix, whose form is motivated by fluctuations in the vacuum energy, which have been shown to be conformal near a light-sheet horizon.  For the experimental measurement of interest in an interferometer, the size of the energy fluctuations is fixed by the area of a surface bounding the volume of spacetime being interrogated by an interferometer.  We compute the interferometer response to these ``geontropic'' scalar-metric fluctuations, and apply our results to current and future interferometer measurements, such as LIGO and the proposed GQuEST experiment.
\end{abstract}

	\maketitle
	\section{Introduction} \label{sec:introduction}
	
	Traditional wisdom in effective field theory (EFT) suggests that quantum fluctuations in the fabric of spacetime should be of the order of $\sim l_p=\sqrt{8\pi G\hbar/c^3}\sim 10^{-34}$ m, where $G$, $\hbar$, $c$, and $l_p$ are the gravitational constant, reduced Planck constant, speed of light, and Planck length respectively. Fluctuations on such small time and length scales are experimentally undetectable. 
 
    It has, however, been recently argued in multiple different contexts that the length scale $L$ of the physical system itself may enter into the observable~\cite{Verlinde_Zurek_2019_1, Verlinde_Zurek_2019_2, Banks_2021, Zurek_2020, Gukov:2022,Verlinde_Zurek_3} (see Ref.~\cite{Zurek_2022} for a summary)
    \begin{equation} \label{eq:length_fluctuation}
        \left\langle\left(\frac{\Delta L}{L}\right)^2\right\rangle
        \sim\frac{l_p}{L}\,,
    \end{equation}
    where $\Delta L$ is the quantum fluctuation of $L$. For example, in Refs.~\cite{Verlinde_Zurek_2019_1, Zurek_2020}, $L$ is the length of interferometer arm in flat spacetime. More generally, $L$ can be the size of a causal diamond in dS, AdS, and flat spacetime \cite{Banks_2021,Verlinde_Zurek_2019_2}. These works argued that the naive EFT reasoning is corrected by long-range correlations in the metric fluctuations--such as are known to occur in holography--which allow the UV fluctuations to accumulate into the infrared.  A physical analogue is Brownian motion (discussed in Ref.~\cite{Zurek_2022}) where the interactions occur at very short distances but become observable on long timescales as the UV effects accumulate.

    While the calculations presented in Refs.~\cite{Verlinde_Zurek_2019_1, Verlinde_Zurek_2019_2, Banks_2021, Zurek_2020, Gukov:2022} are firmly grounded in standard theoretical techniques, such as AdS/CFT, they have not yet provided important, detailed experimental information, such as the power spectral density.  This was the motivation behind the model of Ref.~\cite{Zurek_2020}, to provide a framework that reproduces important behaviors of the UV-complete theory while also allowing to calculate detailed signatures in the infrared.  In the language of the Brownian motion model, while the fluctuations arise from local interactions, the observable is only defined globally. In the language of an interferometer experiment, one cannot measure spacetime fluctuation within a portion of an interferometer arm length, but must wait for a photon to complete a round trip before making a measurement of the global length fluctuation across the entire arm. 
	
    In this work, we continue along the lines of Ref.~\cite{Zurek_2020}, utilizing a scalar field coupled to the metric to model the behavior of the spacetime fluctuations proposed in Refs.~\cite{Verlinde_Zurek_2019_1, Verlinde_Zurek_2019_2, Banks_2021, Zurek_2020, Gukov:2022}. In particular, we propose a model in four dimensions, where the metric appears as a breathing mode of a sphere controlled by a scalar field $\phi$:
	\begin{equation} \label{eq:metric_pix}
		ds^2=-dt^2+(1-\phi)(dr^2+r^2d\Omega^2)\,.
	\end{equation}
    Since $\phi$ effectively controls the area of a spherical surface, it is thus proportional to the entropy of a causal diamond, and may be identified with the dilaton mode studied in Refs.~\cite{Banks_2021,Gukov:2022}.	In the model we consider, $\phi$ is a scalar field whose quantum fluctuations will be characterized by its occupation number, which we label as $\sigma_{\rm pix}$. The subscript denotes ``pixellon'' following the proposal of Ref.~\cite{Zurek_2020}, referring to the pixels of spacetime whose fluctuations the scalar field is modeling. 
    
\begin{figure}[t] 
		\centering
		\includegraphics[width=1\linewidth]{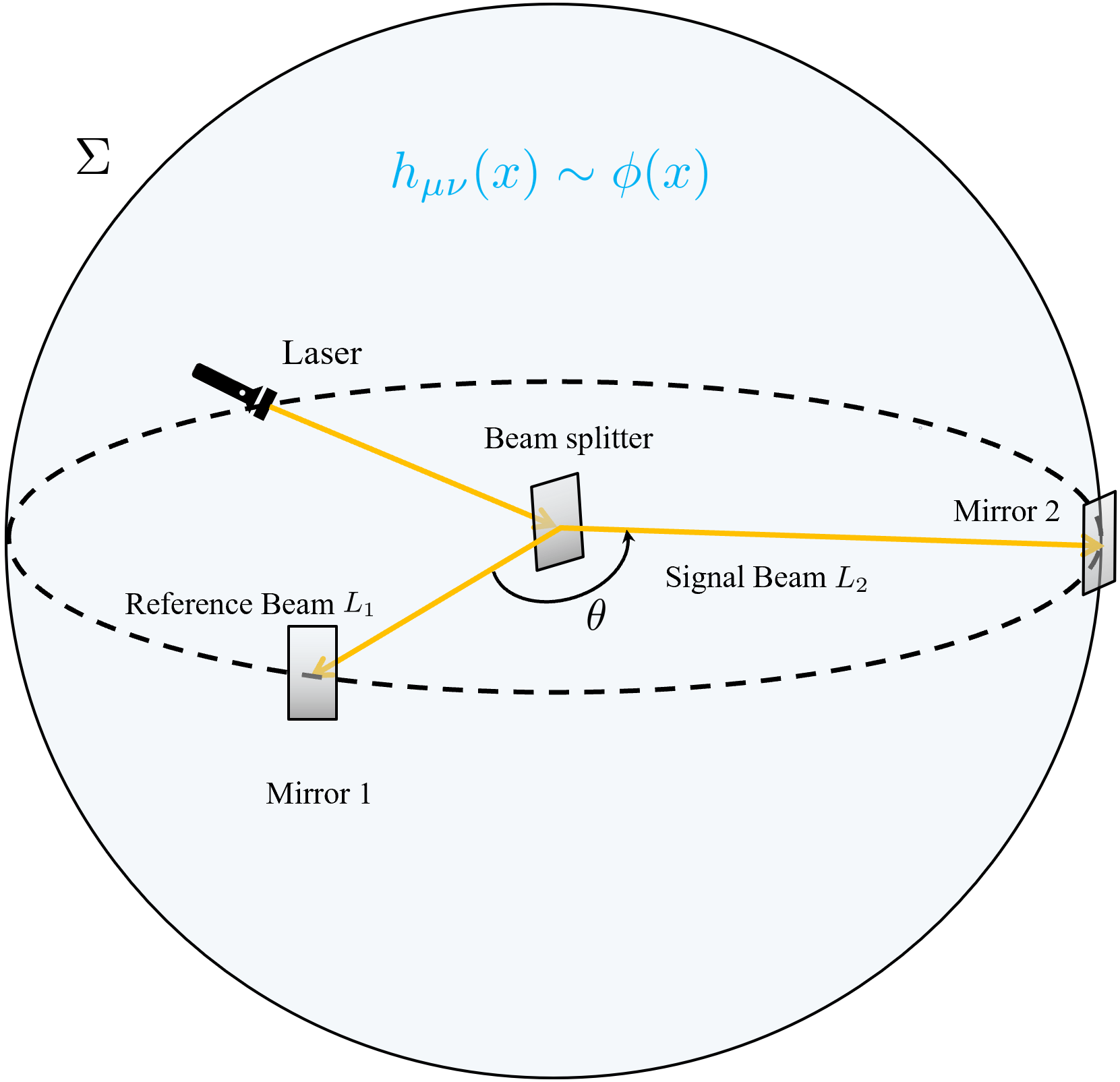}
		\caption{Setup of the interferometer.}
		\label{fig:setup}
	\end{figure}

    In particular, the quantum fluctuations of the scalar, since they couple to the metric, will give rise to fluctuations in the round-trip time for a photon to traverse from mirror to mirror in an interferometer, as depicted in Fig.~\ref{fig:setup}. Similar to Ref.~\cite{Zurek_2020}, our main goal is to compute the gauge invariant interferometer observable arising from the metric Eq.~\eqref{eq:metric_pix}, with $\phi$ being a scalar field having a high occupation number. In contrast to Ref.~\cite{Zurek_2020}, which calculated length fluctuations utilizing the Feynman-Vernon influence functional in a single interferometer arm, we will use only linearized gravity and the QFT of a scalar field with a given occupation number. We will thus be able to extend the previous work in Ref.~\cite{Zurek_2020}, calculating both the power spectral density and angular correlations in the interferometer arms in a manifestly gauge invariant way, checking previous claims made in Ref.~\cite{Verlinde_Zurek_2019_1}, as well as making new predictions. Note that while the model is not yet uniquely derived from first principles in the ultraviolet (utilizing for example shockwave geometry~\cite{Verlinde_Zurek_3}), we will argue below that it is nevertheless well-motivated from first principles.

More specifically, we consider an interferometer with two arms of equal length $L$, {\em i.e.}, with spherical symmetry, and separated by angle $\theta$, as depicted in Fig.~\ref{fig:setup}. We assume that the first arm as the reference beam points in the direction $\mathbf{n}_1$, and the second arm as the signal beam points in the direction $\mathbf{n}_2$. We will find that the observable takes the form:
     \begin{align}
		& \left\langle \frac{\delta T(t_1,{\bf n}_1) \delta T(t_2,{\bf n}_2)}{4L^2} \right\rangle \nonumber\\
		& =\frac{l_p^2}{4L^2}\int_{0}^{L}dr_1\int_{0}^{L}dr_2
	    \int\frac{d^3\mathbf{p}}{(2\pi)^3}     \frac{\sigma_{\pix}(\mathbf{p})}{2\omega(\mathbf{p})}
	    {\cal F}(r_1,r_2,p,\Delta x)\,,
        \label{eq:observable}
    \end{align}
    where $\delta T(t,{\bf n})$ denotes the fluctuation of time delay of light beam sent at time $t-L$ along the direction ${\bf n}$, and $p=(\omega,{\bf p}),~\Delta x=(\Delta t,\Delta {\bf x})$ are four-vectors. The main object of interest in this paper is ${\cal F}(r_1,r_2,p,\Delta x)$, which encapsulates the response of the interferometer gravitationally coupled to the scalar field $\phi$.
    
	The rest of the paper is organized around deriving Eq.~\eqref{eq:observable}. In Sec.~\ref{sec:pixellon}, we review the pixellon scalar field model, with an occupation number $\sigma_{\rm pix}$ motivated in particular by \cite{Zurek_2020}, but also by work demonstrating that the effect of interest is a breathing mode of the horizon \cite{Banks_2021,Gukov:2022}. We then couple this scalar field to the Einstein-Hilbert action and derive its equation of motion. In Sec.~\ref{sec:interferometer_response}, we perform a linearized gravity calculation and derive the observable. In particular, we compute the interferometer response function ${\cal F}(r_1,r_2,p,\Delta x)$ from our specific model. In Sec.~\ref{sec:observations}, we compute the relevant power spectral density and angular correlation from Eq.~\eqref{eq:observable}. We then discuss various existing experimental constraints. Finally, in Sec.~\ref{sec:conclusions}, we conclude. Throughout the paper we will work in units $\hbar=c=k_B=1$ while keeping the gravitational constant $G=l_p^2/(8\pi)$ explicit.

	\section{Scalar Field Quantum Fluctuations in a Causal Diamond} \label{sec:pixellon}

    The main goal of this section is to motivate the form of the scalar occupation number, $\sigma_{\rm pix}$, that will be coupled to the metric. Our discussion here is mostly based on Ref.~\cite{Zurek_2020}, though, as mentioned previously, it is also broadly consistent with the dilaton model presented in Ref.~\cite{Banks_2021,Gukov:2022}.

    The effect of interest, as presented in Refs.~\cite{Verlinde_Zurek_2019_1,Verlinde_Zurek_2019_2} is based on fluctuations in the modular Hamiltonian $K$  
	\begin{equation} \label{eq:modular_K_def}
		 K=\int_{B}T_{\mu\nu}\zeta_{K}^{\mu}dB^{\nu}\,,
	\end{equation}
    where $B$ is some spatial region with a stress tensor $T_{\mu \nu}$, $dB^{\nu}$ is the volume element of $B$ (with $dB^{\nu}$ pointing in the time direction), and $\zeta_{K}^{\mu}$ is the conformal Killing vector of the boost symmetry of $\Sigma$, the entangling surface between $B$ and its complement $\bar{B}$ \cite{Casini_Huerta_Myers_2011, Verlinde_Zurek_2019_2}. One can map $B$ to Rindler space, so $\Sigma$ is also a Rindler horizon. In the context of AdS/CFT, where $T_{\mu \nu}$ is the stress tensor of the boundary CFT, both the vacuum expectation value and the fluctuations of the modular Hamiltonian are known to obey an area law in vacuum \cite{Verlinde_Zurek_2019_2,DeBoer:2018kvc,Nakaguchi:2016zqi}
 	\begin{equation} \label{eq:AdSK} 
		\langle K\rangle
		=\langle \Delta K^2\rangle
		=\frac{A(\Sigma)}{4G}\,,
	\end{equation}
 	where $A(\Sigma)$ is the area of $\Sigma$. 
    One tempting interpretation of this relation is that $\langle K\rangle \equiv {\cal N}$ counts the number of gravitational bits, or pixels, in the system, which is further motivated by the fact that the entanglement entropy $S_{\rm ent} = \langle K \rangle$ is known to hold in a CFT.  The fluctuations of those ${\cal N}$ bits then satisfy ``root-{\cal N}'' statistics:  
    \begin{equation}
        \frac{|\Delta K|}{\langle K \rangle} = \frac{1}{\sqrt{\cal N}}\,, 
        \label{eq:N}
    \end{equation}
    where $|\Delta K| = \sqrt{\langle \Delta K^2 \rangle}$ represents the amplitude of the modular fluctuation.   
  
    While the precise relation $\langle K\rangle
    =\langle \Delta K^2\rangle$ is demonstrated only in the context of AdS/CFT, one can place a Randall-Sundrum brane in the (5-d) bulk of AdS, inducing gravity on the (flat 4-d) RS brane, and show that Eq.~\eqref{eq:AdSK} holds on the 4-d brane \cite{Banks_2021}. The measuring apparatus can then be placed on the flat 4-d brane. Further, as shown in \cite{Solodukhin_1999,Carlip_2012,Banks_2021}, gravity is approximately conformal near the horizon. For an interferometer, the light beams are probing the near-horizon geometry of the spherical entangling surface $\Sigma$ bounding it (shown in Fig.~\ref{fig:setup}), so Ref.~\cite{Banks_2021} argued that the correlator of stress tensor takes the same form as any CFT. Thus, $\langle \Delta K^2\rangle$ follows Eq.~\eqref{eq:AdSK}, {\em i.e.,}
    \begin{eqnarray}
        \langle \Delta K^2 \rangle & \sim & \int d^2{\bf y} d^2 {\bf y}' \frac{dr~dr' r~r'}{((r-r')^2 + ({\bf y}-{\bf y}')^2)^4} \nonumber \\
        &\sim& A \int \frac{dr~dr' r~r'}{(r-r')^6} \sim \frac{A}{\delta^2} \sim \frac{A}{l_p^2}\,,
    \end{eqnarray}
    where $\mathbf{y}$ denotes the transverse directions (corresponding to the coordinates on $\Sigma$), and $G \sim \delta^2$ corresponds to a UV cut-off in the theory at a distance scale $\delta\sim l_p$.  In our case, $r-r' \sim \delta$ corresponds to the distance to the (unperturbed) spherical entangling surface $\Sigma$ in our setup shown in Fig.~\ref{fig:setup}. A similar relation holds for $\langle K \rangle$. More generally, as found in \cite{Srednicki_1993}, an area law for entanglement entropy does not hold only for a CFT but also any massless scalar QFT, which also motivates the scalar model of geoentropic fluctuations in \cite{Zurek_2020} and this work.
  
    The idea of Ref.~\cite{Zurek_2020} was thus to model the gravitational effects of modular fluctuations with a massless scalar field, dubbed a ``pixellon.''      Since pixellons are bosonic scalars, their creation and annihilation operators $(a,a^{\dagger})$ satisfy the usual commutation relation
	\begin{equation} \label{eq:commu}
		\left[a_{\mathbf{p}_1},a_{\mathbf{p}_2}^{\dagger}\right]=
		(2\pi)^3\delta^{(3)}(\mathbf{p}_1-\mathbf{p}_2)\,.
	\end{equation}
    We are interested in modeling the impact of the (fluctuating) effective stress tensor in Eq.~\eqref{eq:stress}.  We will do this by allowing for a non-zero occupation number $\sigma_{\pix}(\mathbf{p})$,
 	\begin{equation} \label{eq:adagger_a}
	    \operatorname{Tr}\left(\rho_{\pix}
		a_{\mathbf{p}_1}^{\dagger}a_{\mathbf{p}_2}\right)
		=(2\pi)^3\sigma_{\pix}(\mathbf{p}_1)
		\delta^{(3)}(\mathbf{p}_1-\mathbf{p}_2)
	\end{equation}
    such that
	\begin{equation} \label{eq:anti_commu_trace}
		\operatorname{Tr}\left(\rho_{\pix}
		\{a_{\mathbf{p}_1},a^{\dagger}_{\mathbf{p}_2}\}\right)
		=(2\pi)^3\left[1+2\sigma_{\pix}(\mathbf{p}_1)\right]
		\delta^{(3)}(\mathbf{p}_1-\mathbf{p}_2)\,.
	\end{equation}
    The occupation number should be consistent with the modular energy fluctuation, Eq.~\eqref{eq:N}, as we will check explicitly at the end of this section. 
    
    The pixellon couples to the metric and sources the stress tensor at second order in perturbations. In general, we can consider a metric of the form
    \begin{equation}
        g_{\mu\nu}
        =\eta_{\mu\nu}+\epsilon h_{\mu\nu} + \epsilon^2 H_{\mu\nu} + ...\,,
    \end{equation}
    where $\epsilon$ is a dimensionless parameter that denotes the order in perturbation theory. The vacuum Einstein Equation (EE) is, parametrically~\footnote{This argument was formulated in private communication with E.~Verlinde in the work leading to Ref.~\cite{Verlinde_Zurek_2019_2}.},
    \begin{equation}
        G_{\mu \nu} = \epsilon \left[\nabla^2 h\right]_{\mu \nu} + \epsilon^2 \left( \left[\nabla^2 H\right]_{\mu \nu} -  l_p^2 T_{\mu \nu}\right) + ... = 0\,,
       \label{eq:vacuumEE}
    \end{equation}
    where the precise form of the equations of motion ({\em e.g.}, numerical prefactors in the time and spatial derivatives) will depend on the precise form of the metric that we consider below, and
    where the effective stress tensor is given by
    \begin{equation} \label{eq:stress}
       T_{\mu \nu} \sim \frac{1}{l_p^2} \left[(\nabla h)^2\right]_{\mu \nu}\,.
    \end{equation}
    
    At leading order in perturbation theory, the metric perturbation $h_{\mu\nu}$ satisfies the vacuum EE having a form
    \begin{equation} \label{eq:EE_first}
        \left[\nabla^2 h\right]_{\mu\nu}=0\,.
    \end{equation}
    However, at second order, the effective stress tensor of $h_{\mu\nu}$ will source a non-zero metric perturbation $H_{\mu\nu}$, {\em i.e.},
    \begin{equation} \label{eq:EE_second}
        \left[\nabla^2 H\right]_{\mu\nu}=l_p^2 T_{\mu \nu}\,.  
    \end{equation}
    One can compute $\langle K\rangle$ from $\langle T_{\mu\nu}\rangle$, but as shown in \cite{Verlinde_Zurek_2019_2}, $\langle K\rangle$ does not gravitate and should be subtracted in the metric equation of motion (similar to a tadpole diagram in QFT). Thus, the vacuum expectation value of this stress tensor vanishes, $\langle T_{\mu\nu}\rangle=0$, consistent with Eqs.~\eqref{eq:stress}-\eqref{eq:EE_first}. In contrast, it is expected to have nonzero fluctuations $\langle\Delta K^2\rangle\sim\langle T_{\alpha \beta} T_{\mu \nu} \rangle \neq 0$, which gravitate and lead to physical observables. 
    
    Although $\langle \Delta K^2\rangle$ is directly related to the {\em vacuum} two-point function of $H_{\mu\nu}$ or four-point function of $h_{\mu\nu}$, the physical observable can be directly computed from the two-point function of $h_{\mu\nu}$ with a nontrivial density-of-states $\sigma_{\pix}$. That is, we are using the language of linearized gravity in this work, while our result captures the nonlinearity in Eq.~\eqref{eq:EE_second} and higher orders via $\sigma_{\pix}$. To compute the fluctuations, we quantize the metric perturbations via the scalar field $\phi$, which, to second order in perturbation theory, leads to a nonzero $\langle \Delta K^2\rangle$, as shown at the end of this section. The major goal of this work is to compute the effects of such quantized metric perturbations on the interferometer depicted in Fig.~\ref{fig:setup}.
   
	More specifically, following Ref.~\cite{Zurek_2020}, we model these energy fluctuations, in the volume of spacetime interrogated with an interferometer, with a thermal density matrix $\rho_{\pix}$, as shown in Eqs.~\eqref{eq:adagger_a}-\eqref{eq:anti_commu_trace}. The motivation for this choice is based on formal work \cite{Casini_Huerta_Myers_2011} showing that the reduced density matrix $\rho_{V}$ of the system $V$ bounded by a sphere $S^{d-1}$ or its casual development $\mathcal{D}$ can be mapped to the thermal density matrix $\rho_{\beta}$ of the hyperbolic spacetime $R\times H^{d-1}$, which foliates $\AdS_{d+1}$, in the asymptotic limit. A similar argument relating the vacuum state of any QFT in a causal diamond to a thermal density matrix can be found in \cite{Jacobson_2016}. 
 
    Thus, following \cite{Zurek_2020}, we are motivated to define a thermal density matrix $\rho_{\pix}$ of pixellons using the definition in \cite{Casini_Huerta_2009},
	\begin{align} \label{eq:density_matrix}
		& \rho_{\pix}
		=\frac{1}{\mathcal{Z}}\exp\left[-\beta\int\frac{d^3\mathbf{p}}{(2\pi)^3}
		(\epsilon_{\mathbf{p}}-\mu)
		a_{\mathbf{p}}^{\dagger}a_{\mathbf{p}}\right]\,, \\
		& \mathcal{Z}
		=\prod_{\mathbf{p}}\frac{1}{1-e^{-\beta(\epsilon_{\mathbf{p}}-\mu)}}\,,
	\end{align}
	where $\epsilon_{\mathbf{p}}$ is the energy of pixellons with momentum $\mathbf{p}$, and $\mu$ is the chemical potential counting background degrees of freedom associated with $\langle K\rangle$ \cite{Zurek_2020}. 

    Furthermore, as in Ref.~\cite{Zurek_2020}, we identify the energy per degree-of-freedom as
	\begin{equation} \label{eq:energy_momentum}
	    \beta(\epsilon_{\mathbf{p}}-\mu) 
	    \equiv \beta\omega({\bf p})\sim\frac{|\Delta K|}{\langle K\rangle}\,.
	\end{equation}
	In four dimensions, according to Eq.~\eqref{eq:AdSK},
    \begin{equation}
        \frac{|\Delta K|}{\langle K\rangle}
        =\frac{1}{\sqrt{\mathcal{N}}}\sim\frac{l_p}{L}\,,
    \end{equation}
    suggesting that the energy fluctuation per degree-of-freedom is set by a ratio of UV and IR length scales. Since $\frac{l_p}{L}\ll 1$, we approximate the occupation number $\sigma(\mathbf{p})$ by
	\begin{equation} \label{eq:dos_approx}
	    \sigma_{\pix}(\mathbf{p})
	    =\frac{1}{e^{\beta\omega(\mathbf{p})}-1}
		\approx\frac{1}{\beta\omega(\mathbf{p})}\,.
	\end{equation}
    More specifically, we identify the IR length scale $1/L \sim \omega(\mathbf{p})$, so we take 
    \begin{equation} \label{eq:dos}
        \sigma_{\pix}(\mathbf{p})
        =\frac{a}{l_p \omega({\bf p})}\,, 
    \end{equation}
    where $a$ is the dimensionless number to be measured in an experiment, or fixed in a UV-complete theory. Here $a = 1/(2\pi)$ corresponds to an inverse temperature $\beta = 2 \pi l_p$, giving a result most closely mirroring Refs.~\cite{Verlinde_Zurek_2019_1,Verlinde_Zurek_2019_2,Zurek_2020} in amplitude. 
    
    Note that $\sigma_{\pix}(\mathbf{p})$ is not Lorentz invariant, but this is to be expected because the measurement of interest via a causal diamond picks out a frame. This is also not contradictory to our statement that we have computed a gauge invariant observable. It is because Lorentz transformations of $\sigma_{\pix}(\mathbf{p})$ are global transformations of background Minkowski spacetime. After the interferometer picks a frame, the interferometer response is independent of how we describe metric perturbations, {\em i.e.}, independent of local coordinate transformations at scale of metric perturbations, which is what gauge invariance usually means in linearized gravity. 
    
    We now derive the dispersion relation for the scalar field from the metric in Eq.~\eqref{eq:metric_pix}.
    We start from the linearized Einstein Hilbert action or Fierz-Pauli action~\cite{Hinterbichler_2012}
	\begin{equation} \label{eq:FP_action}
	\begin{aligned}
		S_{\FP}
		=& \;\frac{1}{2\kappa}\int d^4x\;\sqrt{-g}\;
		h_{\mu\nu}\left(G^{\mu\nu}[h_{\mu\nu}]-\kappa T^{\mu\nu}\right) \\
		=& \;\frac{1}{4\kappa}\int d^{4}x\;\sqrt{-g}\;
		h_{\mu\nu}(\eta^{\mu\nu}\square h-\square h^{\mu\nu} \\
		& \;-2\nabla^{\mu}\nabla^{\nu}h
		+2\nabla_{\rho}\nabla^{\mu}h^{\nu\rho}
		-2\kappa T^{\mu\nu})+\mathcal{O}(h^3)\,,
	\end{aligned}
	\end{equation}
	where $\kappa=8\pi G$. The Fierz-Pauli action can be derived by expanding the full metric $g_{\mu\nu}$ about the Minkowski metric $\eta_{\mu\nu}$, $g_{\mu\nu}=\eta_{\mu\nu}+h_{\mu\nu}$, and keeping the terms quadratic in $h_{\mu\nu}$ in the Einstein Hilbert action \cite{Hinterbichler_2012, Parikh_Wilczek_Zahariade_2020}. Here, $h_{\mu\nu}$ is the metric perturbation associated with the pixellon $\phi$. The terms linear in $h_{\mu\nu}$ are discarded because they can be written as a total derivative \cite{Parikh_Wilczek_Zahariade_2020}. 
	
    Instead of a functional of a general $h_{\mu\nu}$, $S_{\FP}$ in our model is a functional of the metric in Eq.~\eqref{eq:metric_pix} and thus a functional of $\phi$, so the pixellon's action $S_{\pix}[\phi]$ is
	\begin{equation} \label{eq:action_pix_def}
		S_{\pix}[\phi]\equiv S_{\FP}[h^{\pix}_{\mu\nu}[\phi]]\,,\quad
		h_{\mu\nu}^{\pix}\;dx^{\mu}dx^{\nu}=ds^2_{\pix}\,,
	\end{equation}
	which after plugging in Eq.~\eqref{eq:metric_pix} becomes
	\begin{align} \label{eq:action_pix}
		& S_{\pix}[\phi]=\frac{1}{2\kappa}\int d^{4}x\;\sqrt{-g}\;
		\phi\left[\nabla^2-3\partial_t^2\right]\phi
		+\kappa\mathcal{L}_{\interact}[\phi]\,,\nonumber\\
		& \mathcal{L}_{\interact}[\phi]\equiv
		-h_{\mu\nu}^{\pix}[\phi]T^{\mu\nu}\,.
	\end{align}  
	Then the equation of motion (EOM) of $\phi$ is derived by varying $\mathcal{L}_{\pix}$ with respect to $\phi$.
	\begin{equation} \label{eq:EOM_pix}
		\left(\partial_t^2-c_s^2\nabla^2\right)\phi=\frac{\kappa}{c_s^2}
		\frac{\delta\mathcal{L}_{\interact}[\phi]}{\delta\phi}\,,\quad
		c_s\equiv\sqrt{\frac{1}{3}}\,.
	\end{equation}
    
    Following the logic of Eqs.~\eqref{eq:vacuumEE}-\eqref{eq:stress}, to leading order in $\phi$, the right-hand side of Eq.~\eqref{eq:EOM_pix} vanishes. Although Eq.~\eqref{eq:EOM_pix} is source-free, one may find that the effective stress tensor contains linear term in $\phi$, which is a tadpole due to imposing the form of metric in Eq.~\eqref{eq:metric_pix} and can be subtracted off. Eq.~\eqref{eq:EOM_pix} also implies that for the metric in Eq.~\eqref{eq:metric_pix}, $\phi$ needs to have the dispersion relation 
	\begin{equation} \label{eq:dispersion}
	    \omega=c_s|\mathbf{p}|\,,\quad
	    c_s=\sqrt{\frac{1}{3}}
	\end{equation}
	using the expansion $\phi=\int \frac{d^3\mathbf{p}}{(2\pi)^3}\;\phi(\mathbf{p})e^{-i\omega t+i\mathbf{p}\cdot\mathbf{x}}$. It is clear that $\phi$ is a sound mode with the sound speed $c_s=\sqrt{\frac{1}{3}}$. From Eq.~\eqref{eq:action_pix}, we also notice that to canonically normalize $\phi$, we can define $\bar{\phi}$ such that
	\begin{equation} \label{eq:normalization}
		\phi=\sqrt{\kappa}\bar{\phi}=l_p\bar{\phi}\,.
	\end{equation}

	As a consistency check, one can use the metric in Eq.~\eqref{eq:metric_pix} and the occupation number in Eq.~\eqref{eq:dos} to confirm that $\langle \Delta K^2\rangle$ has the same scaling in Eq.~\eqref{eq:AdSK}. Although the physical observable is driven by the two-point function of $\phi$ as we will discuss in Sec.~\ref{sec:interferometer_response}, $\langle \Delta K^2\rangle$ is driven by the four-point function of $\phi$.  One can see this by noting that 
    $K^2 \sim (T_{\mu \nu})^2$ according to Eq.~\eqref{eq:modular_K_def}, while $T_{\mu \nu} \sim \frac{1}{l_p^2}\left[(\nabla \phi)^2\right]_{\mu \nu}$ according to Eq.~\eqref{eq:stress}. In Sec.~\ref{sec:interferometer_response}, we find, utilizing the Ansatz Eq.~\eqref{eq:dos} for the density of states, $\langle \phi^2\rangle \sim \frac{l_p}{L}$ [see Eq.~\eqref{eq:phi1phi2_final}]. Thus, if we identify spatial gradients with the IR length scale $1/L$, we obtain $\langle\Delta K^2\rangle\sim\frac{L^2}{l_p^2}\sim \frac{A}{4G}$, as expected.

	\section{Time Delay in Pixellon Model} \label{sec:interferometer_response}	 
	
    The major goal of this work is to compute an interferometer response to fluctuations in the pixellon model. Instead of using the Feynman-Vernon influence functional approach to compute the mirror's motion, {\em e.g.}, in \cite{Zurek_2020, Parikh_Wilczek_Zahariade_2020, Richardson_Kwon_Gustafson_Hogan_Kamai_McCuller_Meyer_Stoughton_Tomlin_Weiss_2021}, we compute the time delay of a light beam traveling a round trip directly. 
    
    In general, for a metric in the form 
    \begin{equation} \label{eq:metric_general}
		ds^2=-(1-\mathcal{H}_0)dt^2+(1+\mathcal{H}_2)dr^2
		+2\mathcal{H}_1dtdr+\cdots\,,
	\end{equation}
	we need to consider three effects: the shift in the clock rate, mirror motion, and light propagation. As discussed in detail in Appendix~\ref{appendix:time_delay_general}, the shift in the clock's rate only depends on $\mathcal{H}_0$, the mirror motion in the radial direction is affected by $\mathcal{H}_{0,1}$, and the light propagation is determined by all three components $\mathcal{H}_{0,1,2}$.
	
	In Appendix~\ref{appendix:gauge_invariance}, we further show that if we take all of these three effects into consideration and sum up the resulting time delay for both outbound and inbound light, the total time delay $T$ of a round trip is gauge invariant, so $T$ is a physical quantity to measure. In this section, we compute the shift of $T$ due to geoentropic fluctuations and its correlation function using the metric of the pixellon model in Eq.~\eqref{eq:metric_pix}. To calculate time delay in a generic metric like Eq.~\eqref{eq:metric_general}, one can refer to Appendix~\ref{appendix:time_delay_general}.
	
	For the metric in Eq.~\eqref{eq:metric_pix}, the only nonzero component in the $t-r$ sector of the metric is $\mathcal{H}_2$, so we only need to consider light propagation. Then for a light beam sent at time $t-L$ along the direction $\mathbf{n}$, its total time delay $T(t,\mathbf{n})$ of a round trip is completely determined by the pixellon field $\phi$, {\em e.g.},
	\begin{align} \label{eq:time_delay_total}
	    & T(t,\mathbf{n})
	    =2L+\frac{1}{2}\int_{0}^{L}dr\;[\phi(x)+\phi(x')]\,, \nonumber \\
	    & x\equiv(t-L+r, r\mathbf{n})\,,\;
	    x'\equiv(t+L-r, r\mathbf{n})\,.
	\end{align}
    We have chosen the start time to be at $t-L$ such that the time coordinate of $x$ and $x'$ are symmetric about $t$.
     
	Since $\phi$ satisfies the massless free scalar wave equation with the sound speed $c_s=\frac{1}{\sqrt{3}}$ [{\em i.e.}, Eqs.~\eqref{eq:EOM_pix} and \eqref{eq:dispersion}], the quantization for $\phi(x)$ should be
	\begin{equation} \label{eq:phi_quant}
		\begin{aligned}
			\phi(x)
			=& \;l_p\int\frac{d^3\mathbf{p}}{(2\pi)^3}
			\frac{1}{\sqrt{2\omega(\mathbf{p})}}
			\left(a_{\mathbf{p}}e^{ip\cdot x}
			+a_{\mathbf{p}}^{\dagger}e^{-ip\cdot x}\right)\,,
		\end{aligned}
	\end{equation}
	where $l_p$ is to make $\bar{\phi}(x)$ canonically normalized, as discussed in Eq.~\eqref{eq:normalization}. Creation and annihilation operators $a_{\mathbf{p}}$, $a^{\dagger}_{\mathbf{p}}$ satisfy the commutation relation in Eq.~\eqref{eq:commu} with a thermal density matrix $\rho_{\pix}$ defined in Eqs.~\eqref{eq:density_matrix} and \eqref{eq:dos}.

	Let us define $\delta T(t,\mathbf{n})$ to be the correction to the total time delay $T(t,\mathbf{n})$. We write the auto-correlation of $\delta T(t,\mathbf{n})$ as
	\begin{align}\label{eq:corr_def}
		& C(\Delta t,\theta)\equiv
		\left\langle\frac{\delta T(t_1,\mathbf{n}_1)\delta T(t_2,\mathbf{n}_2)}{4L^2}\right\rangle\,, \nonumber\\
		& \Delta t\equiv t_1-t_2\,,\quad
		\theta=\cos^{-1}{(\mathbf{n}_1\cdot\mathbf{n}_2)}\,,
	\end{align}
	and using Eq.~\eqref{eq:time_delay_total}, we obtain
	\begin{equation} \label{eq:dTdT_1}
		\begin{aligned}
			C(\Delta t,\theta)
			=& \;\frac{1}{16L^2}\int_0^{L}dr_1\int_{0}^{L}dr_2 \\
			& \;\langle\left(\phi(x_1)+\phi(x_1')\right)
			\left(\phi(x_2)+\phi(x_2')\right)\rangle\,,
		\end{aligned}
	\end{equation}
	where $\langle \mathcal{O}\rangle$ is a shorthand notation for
	\begin{equation}
		\langle \mathcal{O}\rangle=\Tr(\rho_{\pix}\mathcal{O})\,.
	\end{equation}
	We have assumed that $C(\Delta t,\theta)$ only depends on $\Delta t$, the difference of the time when the two beams are sent, and $\theta$, the angular separation of two arms. We will see that this assumption is true. 
	
	Besides the correlation function in Eq.~\eqref{eq:corr_def}, a more physical correlation function is to first subtract the time delay of the first arm $T(t,\mathbf{n}_1)$ from the time delay of the second arm $T(t,\mathbf{n}_2)$, where two beams are sent at the same time $t$, and then correlate this difference of time delay at different beam-sent time:
	\begin{align}\label{eq:corr_T_def}
		& C_{\mathcal{T}}(\Delta t,\theta)
		\equiv\left\langle\frac{\mathcal{T}(t_1,\theta)\mathcal{T}(t_2,\theta)}{4L^2}\right\rangle\,, \nonumber\\
		& \mathcal{T}(t,\theta)\equiv T(t,\mathbf{n}_2)-T(t,\mathbf{n}_1)
		= \delta T(t,\mathbf{n}_2)-\delta T(t,\mathbf{n}_1)\,,
	\end{align}
	such that
	\begin{equation} \label{eq:tautau}
		\begin{aligned}
			& C_{\mathcal{T}}(\Delta t,\theta) 
			=2\left[C(\Delta t,0)-C(\Delta t,\theta)\right]\,.
		\end{aligned}
	\end{equation}
	Here, we treat the first arm as the reference beam and the second arm as the signal beam. Since the relation between $C(\Delta t,\theta)$ and $C_{\mathcal{T}}(\Delta t,\theta)$ is directly given by Eq.~\eqref{eq:tautau}, we will focus on $C(\Delta t,\theta)$ in our calculations below.
	To compute $C(\Delta t,\theta)$ in Eq.~\eqref{eq:dTdT_1}, we need to first compute the correlation function of $\phi$. Using Eq.~\eqref{eq:phi_quant}, we obtain
	\begin{equation} \label{eq:phi_plus_phi'}
		\begin{aligned}
			\phi(x)+\phi(x')
			=& \;l_p\int\frac{d^3\mathbf{p}}{(2\pi)^3}
			\frac{1}{\sqrt{2\omega(\mathbf{p})}}
			2\cos{\left[\omega(L-r)\right]} \\
			& \;\left(a_{\mathbf{p}}
			e^{-i\omega t+i\mathbf{p}\cdot\mathbf{x}}
			+a_{\mathbf{p}}^{\dagger}
			e^{i\omega t-i\mathbf{p}\cdot\mathbf{x}}\right)\,.
		\end{aligned}
	\end{equation}
	Then we have
	\begin{equation} \label{eq:phi1phi2}
		\begin{aligned}
			& \langle\left(\phi(x_1)+\phi(x_1')\right)
			\left(\phi(x_2)+\phi(x_2')\right)\rangle \\
			=& \;4 l_p^2\int\frac{d^3\mathbf{p}_1}{(2\pi)^3}
			\int\frac{d^3\mathbf{p}_2}{(2\pi)^3}\;
			\frac{1}{\sqrt{4 \omega_1(\mathbf{p}_1)
					\omega_2(\mathbf{p}_2)}} \\
			& \cos{\left[\omega_1(L-r_1)\right]}
			\cos{\left[\omega_2(L-r_2)\right]} \\
			& \;\left[\langle a_{\mathbf{p}_1}a^{\dagger}_{\mathbf{p}_2}\rangle
			e^{-i\left(\omega_1t_1-\omega_2t_2
				-\mathbf{p}_1\cdot\mathbf{x}_1
				+\mathbf{p}_2\cdot\mathbf{x}_2\right)}+c.c.\right]\,,
		\end{aligned}
	\end{equation}
	where we have only kept the term proportional to $a_{\mathbf{p}_1}^{\dagger}a_{\mathbf{p}_2}$ and $a_{\mathbf{p}_1}a^{\dagger}_{\mathbf{p}_2}$ since the other terms are zero. 
	
	To evaluate Eq.~\eqref{eq:phi1phi2}, we need to calculate $\langle a_{\mathbf{p}_1}^{\dagger}a_{\mathbf{p}_2}\rangle$ and $\langle a_{\mathbf{p}_1}a^{\dagger}_{\mathbf{p}_2}\rangle$. The former is given directly by Eq.~\eqref{eq:adagger_a}, $\langle a_{\mathbf{p}_1}^{\dagger}a_{\mathbf{p}_2}\rangle=\Tr{(\rho_{\pix}a_{\mathbf{p}_1}^{\dagger}a_{\mathbf{p}_2})}=(2\pi)^3\sigma_{\pix}(\mathbf{p}_1)\delta^{(3)}(\mathbf{p}_1-\mathbf{p}_2)$. Using both Eq.~\eqref{eq:adagger_a} and the commutation relation in Eq.~\eqref{eq:commu}, we find the latter to be
	\begin{align} \label{eq:a_adagger}
		\langle a_{\mathbf{p}_1}a^{\dagger}_{\mathbf{p}_2}\rangle
		=& \;(2\pi)^3[1+\sigma_{\pix}(\mathbf{p}_1)]
		\delta^{(3)}(\mathbf{p}_1-\mathbf{p}_2) \nonumber\\
		\approx& \;(2\pi)^3\sigma_{\pix}(\mathbf{p}_1)
		\delta^{(3)}(\mathbf{p}_1-\mathbf{p}_2)\,,
	\end{align}
	where we have used $\sigma_{\pix}(\mathbf{p})\gg1$ at the last line. Then,
	\begin{equation} \label{eq:phi1phi2_final}
		\begin{aligned}
			& \langle\left(\phi(x_1)+\phi(x_1')\right)
			\left(\phi(x_2)+\phi(x_2')\right)\rangle \\
			=& \;4 l_p^2
			\int\frac{d^3\mathbf{p}}{(2\pi)^3}
			\frac{\sigma_{\pix}(\mathbf{p})}{2 \omega(\mathbf{p})}
			\cos{\left[\omega(L-r_1)\right]}
			\cos{\left[\omega(L-r_2)\right]} \\
			& \;\left[e^{-i\omega\Delta t+i\mathbf{p}\cdot\Delta\mathbf{x}}+c.c.\right]\,,
		\end{aligned}
	\end{equation}
	where we have defined $\Delta\mathbf{x}\equiv\mathbf{x}_1-\mathbf{x}_2$. Notice that Eq.~\eqref{eq:phi1phi2} is a complex function in general, so we usually need to symmetrize it over $\mathbf{x}_{1,2}$. Due to our approximation in Eq.~\eqref{eq:a_adagger}, Eq.~\eqref{eq:phi1phi2_final} is a real function, so the one after symmetrization over $\mathbf{x}_{1,2}$ is the same as Eq.~\eqref{eq:phi1phi2_final}. For simplicity, we will drop the term $c.c.$ and always assume that a complex conjugate is taken.
	
    Finally, plugging Eq.~\eqref{eq:phi1phi2_final} into Eq.~\eqref{eq:dTdT_1}, we obtain
	\begin{equation} \label{eq:dTdT_2}
		\begin{aligned}
			C(\Delta t,\theta)
			=& \;\frac{l_p^2}{4L^2}
			\int_{0}^{L}dr_1\int_{0}^{L}dr_2
			\int\frac{d^3\mathbf{p}}{(2\pi)^3}
			\frac{\sigma_{\pix}(\mathbf{p})}{2\omega(\mathbf{p})} \\
			& \;\cos{\left[\omega(L-r_1)\right]}
			\cos{\left[\omega(L-r_2)\right]} 
			e^{-i\omega\Delta t+i\mathbf{p}\cdot\Delta\mathbf{x}}\,.
		\end{aligned}
	\end{equation}
    This is our main result, and we will work on applying it to existing interferometer configurations next.

    \section{Observational Signatures and Constraints} \label{sec:observations}

    After plugging $\sigma_{\pix}(\mathbf{p})$ in Eq.~\eqref{eq:dos}, Eq.~\eqref{eq:dTdT_2} is reduced to
	\begin{equation} \label{eq:dTdT_3}
		\begin{aligned}
			C(\Delta t,\theta)
			=& \;\frac{al_p}{8L^2}
			\int_{0}^{L}dr_1\int_{0}^{L}dr_2\;
			\int\frac{d^3\mathbf{p}}{(2\pi)^3}
			\frac{1}{\omega^2(\mathbf{p})} \\
			& \;\cos{\left[\omega(L-r_1)\right]}
			\cos{\left[\omega(L-r_2)\right]}
			e^{-i\omega\Delta t+i\mathbf{p}\cdot\Delta\mathbf{x}}\,.
		\end{aligned}
	\end{equation}
	In the next two subsections, we will study the power spectral density and angular correlation of Eq.~\eqref{eq:dTdT_3} in more detail. 

	\subsection{Power spectral density}	\label{sec:PSD}
	
\begin{figure}[t] 
		\centering
		\includegraphics[width=1\linewidth]{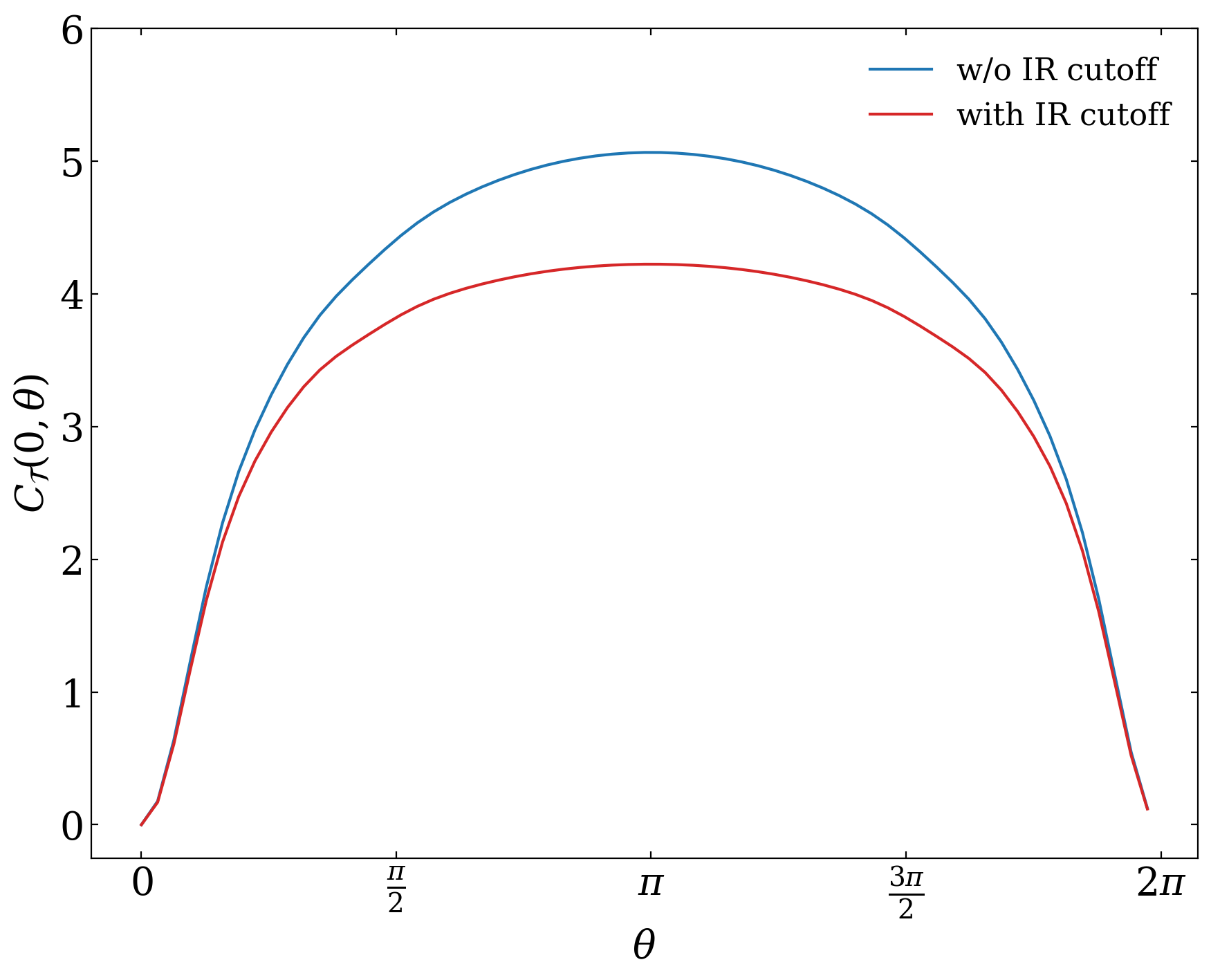}
		\caption{Equal-time correlation function $C_{\mathcal{T}}(0,\theta)$ [{\em i.e.}, Eq.~\eqref{eq:corr_T_def}] of the pixellon model without IR cutoff in Eq.~\eqref{eq:dTdT_3} (blue) and with an IR cutoff in Eq.~\eqref{eq:dTdT_IRCutoff} (red), where both curves are normalized by $\frac{8\pi^2c_s^2L}{al_p}$.}
		\label{fig:corr_equal_time}
	\end{figure}

    We first study the power spectral density implied by Eq.~\eqref{eq:dTdT_3}. Carrying out the angular part of the momentum integral in Eq.~\eqref{eq:dTdT_3}, we have
	\begin{equation} \label{eq:corr}
	\begin{aligned}
		C(\Delta t,\theta)
		=& \;\frac{al_p}{32\pi^2c_s^2L^2}
		\int_{0}^{L}dr_1\int_{0}^{L}dr_2\;\int_{0}^{\infty} d|\mathbf{p}| \\
		& \;\cos{\left[\omega(L-r_1)\right]}
		\cos{\left[\omega(L-r_2)\right]} \\
		& \;e^{-i\omega\Delta t}
		\int_{0}^{\pi}d\vartheta\;\sin{\vartheta}
		e^{i|\mathbf{p}||\Delta\mathbf{x}|\cos{\vartheta}} \\
		=& \;\frac{al_p}{16\pi^2c_s^2L^2}
		\int_{0}^{L}dr_1\int_{0}^{L}dr_2\int_{0}^{\infty} d\omega\; \\
		& \;\cos{\left[\omega(L-r_1)\right]}
		\cos{\left[\omega(L-r_2)\right]} \\
		& \frac{\sin\left[\omega\mathcal{D}(r_1,r_2,\theta)
			/c_s\right]}{\omega\mathcal{D}(r_1,r_2,\theta)} e^{-i\omega\Delta t} \,,
	\end{aligned}
	\end{equation}
	where we have defined
	\begin{equation}
		\mathcal{D}(r_1,r_2,\theta)
		\equiv|\Delta\mathbf{x}|
		=\sqrt{r_1^2+r_2^2-2r_1r_2\cos{\theta}}\,.
	\end{equation}
	The additional factor of $\frac{1}{c_s^2}$ in Eq.~\eqref{eq:corr} comes from using the dispersion relation in Eq.~\eqref{eq:dispersion}.
	$C_{\mathcal{T}}(\Delta t,\theta)$ is directly given by plugging Eq.~\eqref{eq:corr} into Eq.~\eqref{eq:tautau}. In Fig.~\ref{fig:corr_equal_time}, we have plotted $C_{\mathcal{T}}(\Delta t,\theta)$ over the separation angle $\theta$ of the interferometer for $\Delta t=0$.  Notice the signal is maximal when the interferometer arms are back-to-back.
	
\begin{figure*}[t] 
		\centering
		\includegraphics[width=1\linewidth]{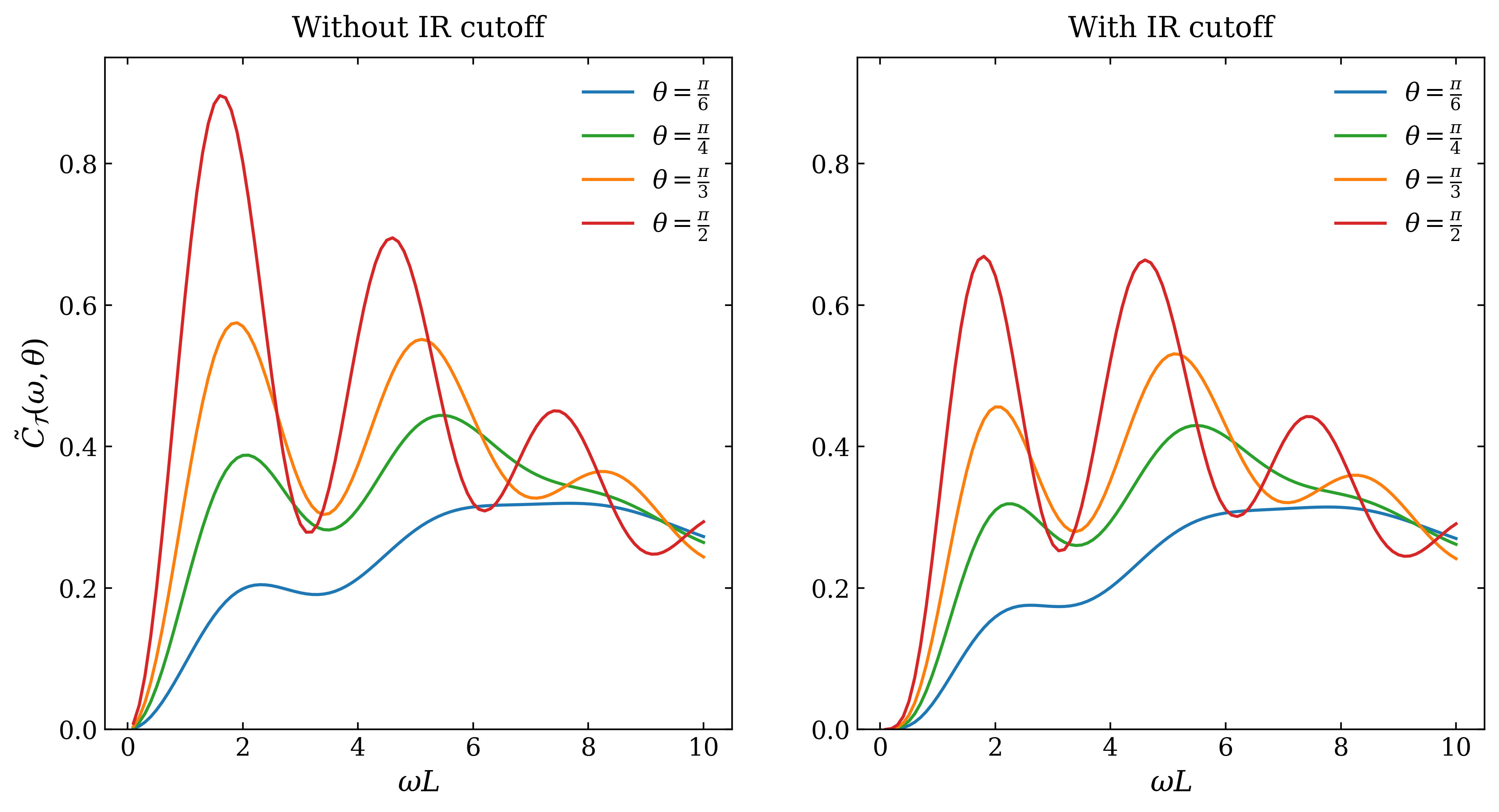}
		\caption{Power spectral density $\tilde{C}_{\mathcal{T}}(\omega,\theta)$ [{\em i.e.}, Eq.~\eqref{eq:C_T_psd}] of the pixellon model without IR cutoff in Eq.~\eqref{eq:C_psd} (left) and with an IR cutoff in Eq.~\eqref{eq:C_psd_cutoff} (right), where all the curves are normalized by $\frac{8\pi c_s^2}{al_p}$. }
		\label{fig:PSD}
	\end{figure*}

	Performing a Fourier transform of $C(\Delta t,\theta)$ with respect to $\Delta t$, we obtain the two-sided power spectral density $\tilde{C}(\omega,\theta)$ to be
	\begin{equation} \label{eq:C_psd}
	\begin{aligned}
		\tilde{C}(\omega,\theta)
		=& \;\int_{-\infty}^{\infty}dt\;e^{-i\omega t}C(t,\theta) \\
		=& \;\frac{al_p}{8\pi c_s^2L^2}
		\int_{0}^{L}dr_1\int_{0}^{L}dr_2 
		\frac{\sin\left[\omega\mathcal{D}(r_1,r_2,\theta)
			/c_s\right]}{\omega\mathcal{D}(r_1,r_2,\theta)} \\
		& \;\cos{\left[\omega(L-r_1)\right]}
		\cos{\left[\omega(L-r_2)\right]}\,.
	\end{aligned}
	\end{equation}
	To evaluate the power spectral density of
	$C_{\mathcal{T}}(\Delta t,\theta)$, we can put Eq.~\eqref{eq:C_psd} into Eq.~\eqref{eq:tautau} such that its power spectral density $\tilde{C}_{\mathcal{T}}(\omega,\theta)$ is
	\begin{equation} \label{eq:C_T_psd}
		\tilde{C}_{\mathcal{T}}(\omega,\theta)=2[\tilde{C}(\omega,0)-\tilde{C}(\omega,\theta)]\,.
	\end{equation}
	In Fig.~\ref{fig:PSD}, we have plotted Eq.~\eqref{eq:C_T_psd} over $\omega L$ for several different separation angles $\theta$ of the interferometer. 
	
	In the limit $\omega\rightarrow0$, Eqs.~\eqref{eq:C_psd}-\eqref{eq:C_T_psd} reduce to
	\begin{align}
		& \tilde{C}(\omega,\theta)
		=\frac{al_p}{8\pi c_s^3}+\mathcal{O}(\omega^2L^2) \,,
		\label{eq:PSD_IR}\\
		& \tilde{C}_{\mathcal{T}}(\omega,\theta)
		=\frac{al_p}{48\pi c_s^5}\omega^2L^2(1-\cos{\theta})
		+\mathcal{O}(\omega^4L^4)\,.
		\label{eq:PSD_diff_IR}
	\end{align}
	A major feature of $\tilde{C}(\omega,\theta)$ at low frequencies is that it is flat in frequency, corresponding to the spectrum of white noise. This feature is consistent with the ``random walk intuition" of holographic effects in \cite{Zurek_2022}, as well as the random walk models in \cite{Amelino_Camelia_1999, Diosi_1985}. On the other hand, although $\tilde{C}(\omega,\theta)$ is independent of $\omega$ at low frequency, $\tilde{C}_{\mathcal{T}}(\omega,\theta)$ is quadratic in $\omega$. It is because, as one can directly observe from Eq.~\eqref{eq:PSD_IR}, the leading order term of $\tilde{C}(\omega,\theta)$ at low frequency is angle-independent. Thus, when subtracting the time delay of the first arm from the second arm, this leading order term cancels out, and the next order term, which is quadratic in $\omega$ and has a nontrivial angular dependence, contributes to $\tilde{C}_{\mathcal{T}}(\omega,\theta)$. 
	
	In Eqs.~\eqref{eq:PSD_IR}-\eqref{eq:PSD_diff_IR}, there are also additional factors of $\frac{1}{c_s}$ from the expansion of $\sin\left[\omega\mathcal{D}(r_1,r_2,\theta)/c_s\right]$ in Eq.~\eqref{eq:C_psd}. Since the leading order term in the expansion of  $\sin\left[\omega\mathcal{D}(r_1,r_2,\theta)/c_s\right]$ is linear in its argument, it contributes an additional factor of $\frac{1}{c_s}$ to $\tilde{C}(\omega,\theta)$ in Eq.~\eqref{eq:PSD_IR}. On the other hand, as we explained above, this leading order term is angle-independent, so the next order term, which is cubic in its argument, contributes an additional factor of $\frac{1}{c_s^3}$ to $\tilde{C}_{\mathcal{T}}(\omega,\theta)$ in Eq.~\eqref{eq:PSD_diff_IR}.
	
	One last observation from Eqs.~\eqref{eq:PSD_IR}-\eqref{eq:PSD_diff_IR} is that both $\tilde{C}(\omega,\theta)$ and $\tilde{C}_{\mathcal{T}}(\omega,\theta)$ are regular in low frequency. In \cite{Verlinde_Zurek_2019_1}, an IR regulator at the scale of $\sim\frac{1}{L^2}$ was added to the 2D Laplacian on the sphere to regulate the angular correlation function as we will discuss in Sec.~\ref{sec:angular_correlation}. To perform an analogous calculation and take into account other IR effects, such as information loss due to soft graviton loss, we will apply the procedures in this section to the pixellon model with an IR cutoff at the same scale as in \cite{Verlinde_Zurek_2019_1} in Sec.~\ref{sec:IR_cutoff}. 
	
	\subsection{Angular correlation} \label{sec:angular_correlation}

    We now study the angular correlation implied by Eq.~\eqref{eq:dTdT_3}. It will be convenient to first decompose Eq.~\eqref{eq:dTdT_3} into spherical harmonics and spherical Bessel functions. Using 
	\begin{equation} \label{eq:plane_wave_decomp}
		e^{i\mathbf{p}\cdot\mathbf{r}}=\sum_{\ell=0}^{\infty}
		i^{l}(2\ell+1)j_{\ell}(|\mathbf{p}|r)
		P_{\ell}(\cos\theta)\,,\quad
		\theta=\hat{\mathbf{p}}\cdot\hat{\mathbf{r}}\,,
	\end{equation}
	and the addition theorem
	\begin{equation}
		P_{\ell}(\hat{\mathbf{p}}\cdot\mathbf{n})
		=\frac{4\pi}{2\ell+1}\sum_{m}
		Y^{\ell m*}(\hat{\mathbf{p}})Y^{\ell m}(\mathbf{n})\,,
	\end{equation}
	we obtain
	\begin{equation}
		\begin{aligned}
			e^{i\mathbf{p}\cdot(\mathbf{x}_1-\mathbf{x}_2)}
			=& \;\sum_{\ell_1,m_1,\ell_2,m_2}16\pi^2i^{\ell_1}(-i)^{\ell_2}
			j_{\ell_1}(|\mathbf{p}|r_1)j_{\ell_2}(|\mathbf{p}|r_2) \\
			& \;Y^{\ell_1m_1*}(\hat{\mathbf{p}})Y^{\ell_2m_2}(\hat{\mathbf{p}})
			Y^{\ell_1m_1}(\mathbf{n}_1)Y^{\ell_2m_2*}(\mathbf{n}_2)\,.
		\end{aligned}
	\end{equation}
	Using $\int d\Omega\;Y^{\ell_1m_1*}(\hat{\mathbf{p}})
	Y^{\ell_2m_2}(\hat{\mathbf{p}})=\delta^{\ell_1\ell_2}\delta^{m_1m_2}$, we can integrate out all the angular dependence of $\mathbf{p}$, so
	\begin{equation} \label{eq:dTdT_4}
		\begin{aligned}
			C(\Delta t,\theta)
			=& \;\frac{al_p}{4\pi c_s^3L^2}\sum_{\ell,m}
			\int_{0}^{L}dr_1\int_{0}^{L}dr_2\int_{0}^{\infty} d\omega \\
			& \;\cos{\left[\omega(L-r_1)\right]}
		    \cos{\left[\omega(L-r_2)\right]} \\
			& \;j_{\ell}(\omega r_1/c_s)j_{\ell}(\omega r_2/c_s) \\
			& \;Y^{\ell m}(\vartheta_1,\varphi_1)
			Y^{\ell m*}(\vartheta_2,\varphi_2)
			e^{-i\omega\Delta t}\,,
		\end{aligned}
	\end{equation}
	where we have an additional factor of $\frac{1}{c_s^3}$ from replacing $\mathbf{p}$ with $\omega$ using Eq.~\eqref{eq:dispersion}. If we define the amplitude of each $(\ell,m)$ mode of the integrand to be
	\begin{equation} \label{eq:A_lm_def}
		\begin{aligned}
			A_{\ell m}(\Delta t,\omega,r_1,r_2)
			\equiv& \;\cos{\left[\omega(L-r_1)\right]}
		    \cos{\left[\omega(L-r_2)\right]} \\
			& \;j_{\ell}(\omega r_1/c_s)j_{\ell}(\omega r_2/c_s)
			e^{-i\omega\Delta t}\,,
		\end{aligned}	
	\end{equation}
	Eq.~\eqref{eq:dTdT_4} can be more compactly written as
	\begin{equation} \label{eq:dTdT_simplified}
		\begin{aligned}
			C(\Delta t,\theta)
			=& \;\frac{al_p}{4\pi c_s^3L^2}\sum_{\ell,m}
			\int_{0}^{L}dr_1\int_{0}^{L}dr_2\int_{0}^{\infty} d\omega \\
			& \;A_{\ell m}(\Delta t,\omega,r_1,r_2) 
			Y^{\ell m}(\vartheta_1,\varphi_1)Y^{\ell m*}(\vartheta_2,\varphi_2)\,.
		\end{aligned}
	\end{equation}

\begin{figure}[t]
		\centering
		\includegraphics[width=1\linewidth]{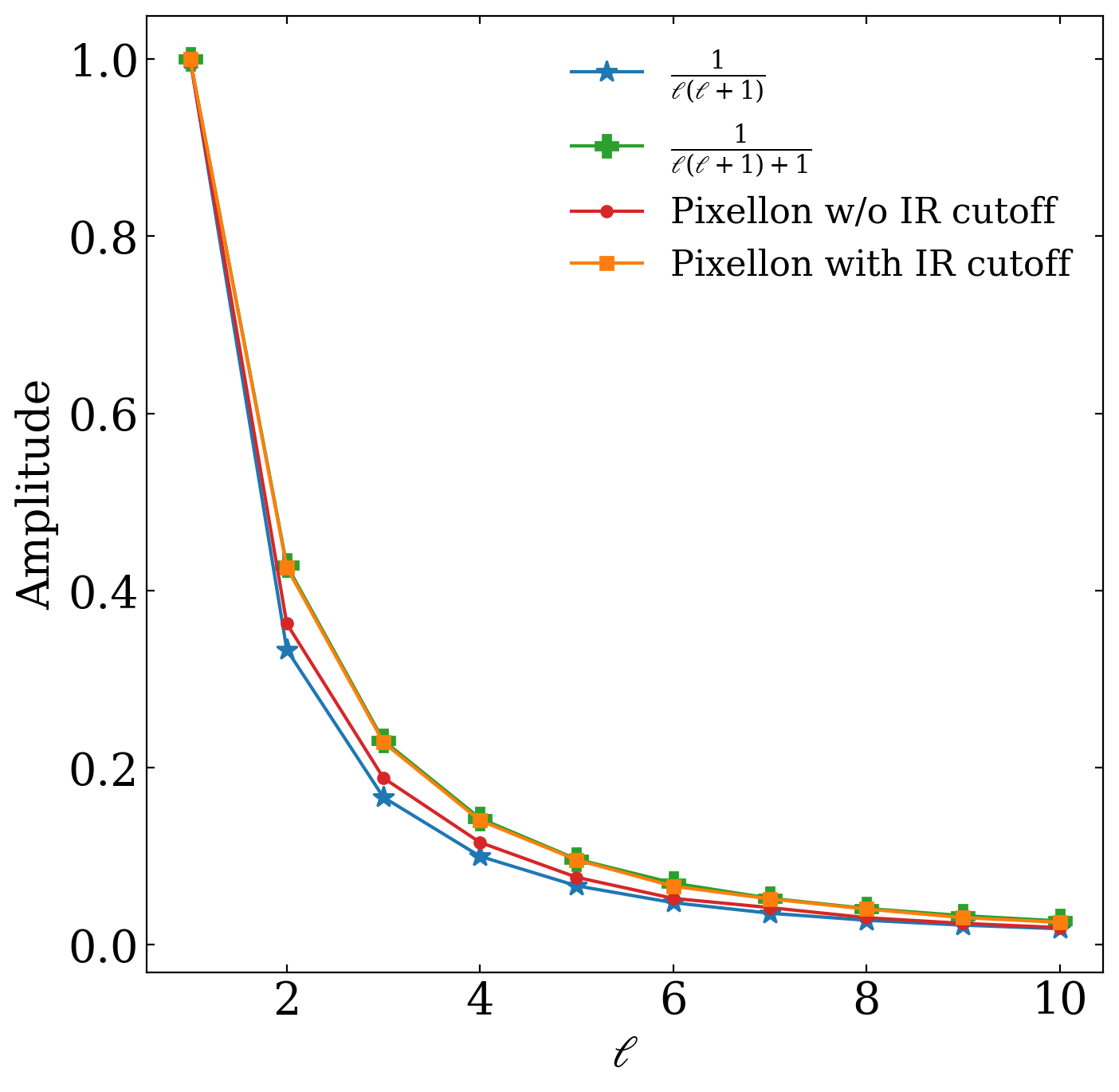}
		\caption{The amplitude of each $(\ell,m)$ mode of the equal-time correlation function $C(0,\theta)$ decomposed into spherical harmonics. The blue and green lines correspond to the amplitude in \cite{Verlinde_Zurek_2019_1} [{\em i.e.}, Eq.~\eqref{eq:angular_pattern_VZ1}] without and with an IR regulator, respectively. The red and orange lines correspond to $c_{\ell m}$ [{\em i.e.}, Eq.~\eqref{eq:c_lm}] of the pixellon model without IR cutoff in Eq.~\eqref{eq:A_lm_def} and with an IR cutoff in Eq.~\eqref{eq:A_lm_cutoff}, respectively. We have normalized the amplitude of each mode by the amplitude of the mode $\ell=1$.}
		\label{fig:angular_decomp} 
	\end{figure}

	Let us first look at the equal-time correlator by setting $\Delta t=0$. The amplitude $c_{\ell m}$ of each $(\ell,m)$ mode of $C(0,\theta)$ is then given by integrating $A_{\ell m}(0,\omega,r_1,r_2)$ over $\omega$ and $r_{1,2}$ as indicated by Eq.~\eqref{eq:dTdT_simplified}, {\em i.e.},
	\begin{equation} \label{eq:c_lm}
        c_{\ell m}=\frac{al_p}{4\pi c_s^3L^2}
        \int_0^L dr_1 \int_0^L dr_2 \int_{0}^{\infty} d\omega\;
        A_{\ell m}(0,\omega,r_1,r_2)\,.
    \end{equation}
	Since these integrals are hard to evaluate analytically, we have plotted the numerical result in Fig.~\ref{fig:angular_decomp}. In Fig.~\ref{fig:angular_decomp}, we have only plotted the modes starting from $\ell=1$ since the $\ell=0$ mode, which is angle-independent, is cancelled out in $C_{\mathcal{T}}(\Delta t, \theta)$ as explained in the previous section.
	
    In Fig.~\ref{fig:angular_decomp}, we have also shown the amplitude of each $(\ell,m)$ mode found in Ref.~\cite{Verlinde_Zurek_2019_1}. They argued that the angular part of $C(0,\theta)$ should be described by the Green's function of the 2D Laplacian on the sphere with an additional IR regulator at the scale of $\frac{1}{L^2}$. After decomposing the Green's function into spherical harmonics, one obtains
	\begin{equation} \label{eq:angular_pattern_VZ1}
		C(0, \theta)
		\propto\sum_{\ell,m}\frac{Y^{\ell m}(\vartheta_1,\varphi_1)
		Y^{\ell m*}(\vartheta_2,\varphi_2)}{\ell(\ell+1)+1}\,.
	\end{equation}
    Excellent agreement between the pixellon model and the expectation of Ref.~\cite{Verlinde_Zurek_2019_1} is observed.

	As mentioned in Sec.~\ref{sec:PSD}, both $\tilde{C}(\omega,\theta)$ and $\tilde{C}_{\mathcal{T}}(\omega,\theta)$ in this work are regular when $\omega\rightarrow0$, even without an IR regulator, {\em e.g.}, Eqs.~\eqref{eq:PSD_IR}-\eqref{eq:PSD_diff_IR}. However, it will still be interesting to study the pixellon model with an IR cutoff due to IR effects from the physical size of the interferometer. We will consider the case with an IR cutoff in Sec.~\ref{sec:IR_cutoff}, but in this section, we first consider only the model without an IR cutoff. Thus, when comparing Eq.~\eqref{eq:c_lm} to Ref.~\cite{Verlinde_Zurek_2019_1}, we drop the additional $1$ in the denominator of Eq.~\eqref{eq:angular_pattern_VZ1}, which appears due to the insertion of an IR regulator. In this case, the amplitude of each $(\ell,m)$ mode becomes $\frac{1}{\ell(\ell+1)}$. In Fig.~\ref{fig:angular_decomp}, one can observe that the angular correlation in this work is very close to the one in \cite{Verlinde_Zurek_2019_1} without the IR regulator. Note that one also observes the same angular dependence in the shockwave geometry ({\em e.g.}, see Refs.~\cite{tHooft_1987,Dray_tHooft_1985,tHooft_1996,Verlinde_Zurek_3}), a connection we would like to study further in our future work.
	
\begin{figure*}[t]
		\centering
		\includegraphics[width=1\linewidth]{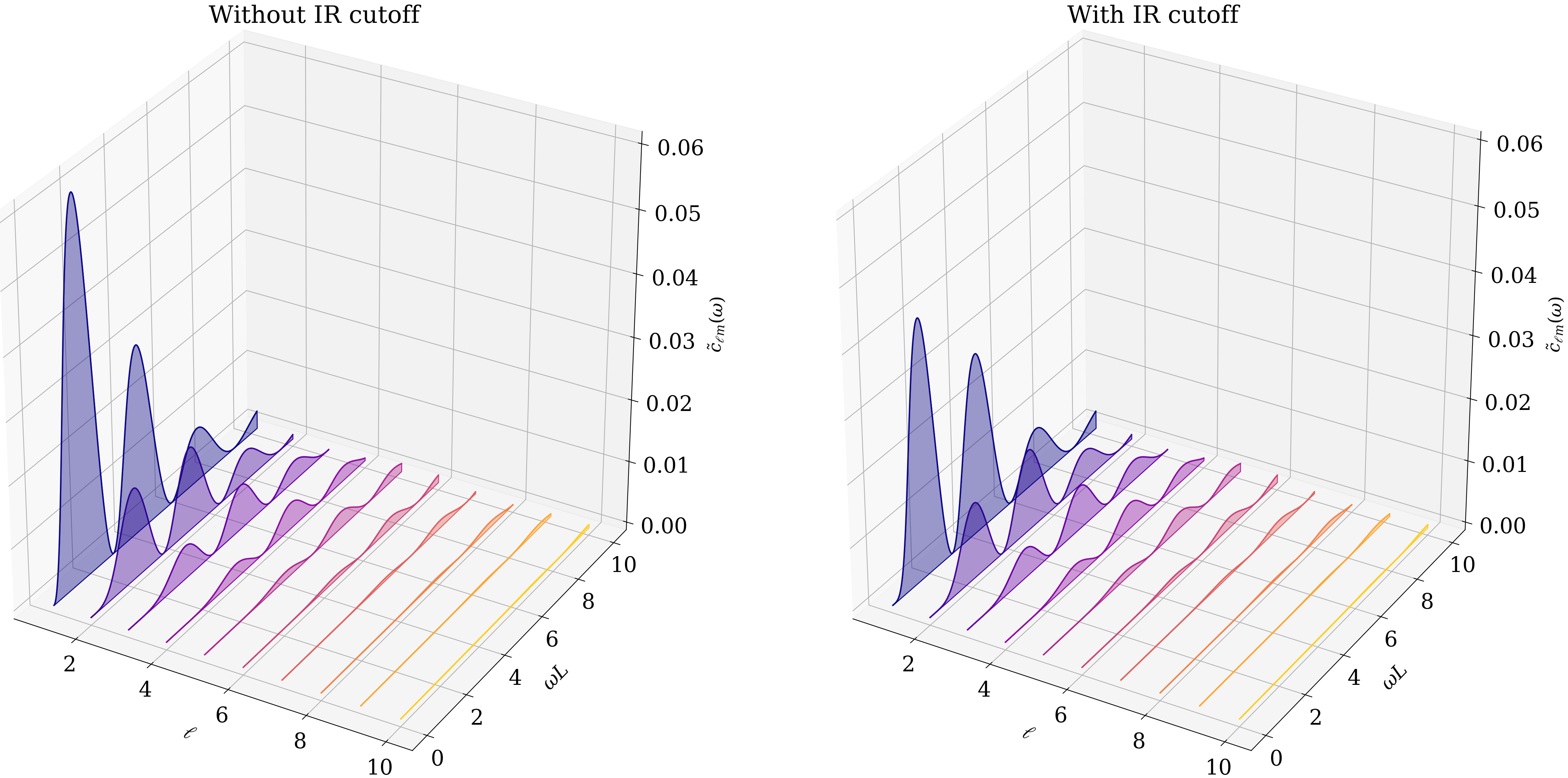}
		\caption{The amplitude $\tilde{c}_{\ell m}(\omega)$ [{\em i.e.}, Eq.~\eqref{eq:tilde_c_lm}] of each $(\ell,m)$ mode of the power spectral density $\tilde{C}(\omega,\theta)$ decomposed into spherical harmonics. The left and right panels are for the pixellon model without IR cutoff in Eq.~\eqref{eq:A_lm_def} and with an IR cutoff in Eq.~\eqref{eq:A_lm_cutoff}, respectively. We have dropped the overall factor $\frac{al_p}{2c_s^3}$ in both plots.}
		\label{fig:angular_decomp_3D} 
	\end{figure*}

	One might also be interested in the amplitude $\tilde{c}_{\ell m}(\omega)$ of each $(\ell,m)$ mode of the power spectral density $\tilde{C}(\omega,\theta)$. Performing a Fourier transform of $C(\Delta t,\theta)$ in Eq.~\eqref{eq:dTdT_simplified} and thus a Fourier transform of $A_{\ell m}(\Delta t,\omega,r_1,r_2)$ in Eq.~\eqref{eq:A_lm_def}, we obtain
	\begin{equation} \label{eq:tilde_c_lm}
	    \tilde{c}_{\ell m}(\omega)=\frac{al_p}{2 c_s^3L^2}
        \int_0^L dr_1 \int_0^L dr_2\; A_{\ell m}(0,\omega,r_1,r_2)\,.
	\end{equation}
	We have plotted $\tilde{c}_{\ell m}(\omega)$ starting from $\ell=1$ in Fig.~\ref{fig:angular_decomp_3D}.
	
	To determine an analytical representation of the amplitude of each $(\ell,m)$ mode, one can also look at $A_{\ell m}(0,\omega,r_1,r_2)$ at the end points $r_1=r_2=L$. If we integrate $A_{\ell m}(0,\omega,L,L)$ over $\omega$, we find the amplitude of each $(\ell,m)$ mode at end points to be
	\begin{equation} \label{eq:A_l_endpoints}
		L\int_{0}^{\infty} d\omega\; A_{\ell m}(0,\omega,L,L)=\frac{\pi c_s}{2(2\ell+1)}\,,	
	\end{equation}
	which is the major contribution to $c_{\ell m}$ plotted in Fig.~\ref{fig:angular_decomp}. Although Eq.~\eqref{eq:A_l_endpoints} decreases more slowly than Eq.~\eqref{eq:angular_pattern_VZ1} over $\ell$, we have additional suppression due to, for example, the factors of $\cos{\left[\omega(L-r_{1,2})\right]}$ in Eq.~\eqref{eq:A_lm_def} when integrating $A_{\ell m}(0,\omega,r_1,r_2)$ over $\omega$ and $r_{1,2}$, so the total amplitude in Eq.~\eqref{eq:c_lm} is very close to Eq.~\eqref{eq:angular_pattern_VZ1} without the IR regulator.

    \subsection{IR cutoff} \label{sec:IR_cutoff}
    
    In this section, we apply the calculations in the previous two sections to the pixellon model with an IR cutoff. As discussed above, although both $\tilde{C}(\omega,\theta)$ and $\tilde{C}_{\mathcal{T}}(\omega,\theta)$ are regular in the IR, we still expect an explicit IR cut-off to enter the calculation because of the finite size of the interferometer.  We will also find that adding an IR cut-off gives a better agreement with the angular correlation of Eq.~\eqref{eq:angular_pattern_VZ1}. For this reason, we place an IR cutoff at a scale $\sim\frac{1}{L^2}$, similar to \cite{Verlinde_Zurek_2019_1}, into Eq.~\eqref{eq:dTdT_3}, {\em e.g.},
    \begin{equation} \label{eq:dTdT_IRCutoff}
		\begin{aligned}
			C(\Delta t,\theta)
			=& \;\frac{al_p}{8L^2}
			\int_{0}^{L}dr_1\int_{0}^{L}dr_2\;
			\int\frac{d^3\mathbf{p}}{(2\pi)^3}
			\frac{1}{\omega^2(\mathbf{p})+\frac{1}{L^2}} \\
			& \;\cos{\left[\omega(L-r_1)\right]}
			\cos{\left[\omega(L-r_2)\right]}
			e^{-i\omega\Delta t+i\mathbf{p}\cdot\Delta\mathbf{x}}\,.
		\end{aligned}
	\end{equation}
	
	Following the same procedure in Sec.~\ref{sec:PSD}, we find that the power spectral density $\tilde{C}(\omega,\theta)$ in Eq.~\eqref{eq:C_psd} is modulated by an additional factor in $\omega$ and $L$, {\em i.e.},
	\begin{equation} \label{eq:C_psd_cutoff}
		\tilde{C}(\omega,\theta)
		\rightarrow\left(\frac{\omega^2}{\omega^2+\frac{1}{L^2}}\right)
		\tilde{C}(\omega,\theta)\,,
	\end{equation}
    while $\tilde{C}_{\mathcal{T}}(\omega,\theta)$ is still given by Eq.~\eqref{eq:C_T_psd}. $C_{\mathcal{T}}(0,\theta)$ and $\tilde{C}_{\mathcal{T}}(\omega,\theta)$ with this IR cutoff are shown in Figs.~\ref{fig:corr_equal_time} and \ref{fig:PSD}, respectively. 
    
    One major effect of the IR cutoff is that the amplitude of $\tilde{C}(\omega,\theta)$ is suppressed at low frequency due to the modulation factor in Eq.~\eqref{eq:C_psd_cutoff}, as one can directly observe in Fig.~\ref{fig:PSD}. For the same reason, the overall amplitude of  $C_{\mathcal{T}}(\Delta t,\theta)$ in the case with an IR cutoff is smaller than the one without IR cutoff as depicted in Fig.~\ref{fig:corr_equal_time}. As frequency increases, the modulation factor goes to $1$, so the amplitude of  $\tilde{C}(\omega,\theta)$ in these two cases becomes nearly identical. In addition, as the separation angle $\theta$ decreases, the difference between these two cases also becomes smaller since interferometers with smaller $\theta$ are more sensitive to higher $\ell$ modes, which have higher characteristic frequency, and thus are less sensitive to the IR cutoff. 
    
    One can also determine the suppression factor due to the IR cutoff as $\omega\rightarrow0$ by expanding Eq.~\eqref{eq:C_psd_cutoff}, {\em e.g.},
    \begin{align}
		& \tilde{C}(\omega,\theta)
		=\frac{al_p}{8\pi c_s^3}\omega^2L^2+\mathcal{O}(\omega^4L^4) \,,
		\label{eq:PSD_IR_cutoff}\\
		& \tilde{C}_{\mathcal{T}}(\omega,\theta)
		=\frac{al_p}{48\pi c_s^5}\omega^4L^4(1-\cos{\theta})
		+\mathcal{O}(\omega^6L^6)\,.
		\label{eq:PSD_diff_IR_cutoff}
	\end{align}
	The IR behaviors of both $\tilde{C}(\omega,\theta)$ and $\tilde{C}_{\mathcal{T}}(\omega,\theta)$ above are very different from the case without an IR cutoff in Eq.~\eqref{eq:PSD_IR}-\eqref{eq:PSD_diff_IR} due to the additional factor of $\omega^2L^2$ contributed by the modulation factor in Eq.~\eqref{eq:C_psd_cutoff}. For this reason, one has to be cautious when constraining our model using detectors with peak sensitivity at low frequency, such as LIGO, as discussed in Sec.~\ref{sec:observations}.
    
	For the angular correlation, after decomposing Eq.~\eqref{eq:dTdT_IRCutoff} into spherical harmonics, we find that the amplitudes $c_{\ell m}$ and $\tilde{c}_{\ell m}(\omega)$ of each $(\ell,m)$ mode of $C(0,\theta)$ and $\tilde{C}(\omega,\theta)$ are given by Eqs.~\eqref{eq:c_lm} and \eqref{eq:tilde_c_lm}, respectively, but $A_{\ell m}(\Delta t,\omega,r_1,r_2)$ is modulated by the same factor in Eq.~\eqref{eq:C_psd_cutoff}, {\em i.e.},
	\begin{equation} \label{eq:A_lm_cutoff}
	    A_{\ell m}(\Delta t,\omega,r_1,r_2)
	    \rightarrow\left(\frac{\omega^2}{\omega^2+\frac{1}{L^2}}\right)
	    A_{\ell m}(\Delta t,\omega,r_1,r_2)\,.
	\end{equation}
	We show both $c_{\ell m}$ and $\tilde{c}_{\ell m}(\omega)$ with the IR cutoff in Figs.~\ref{fig:angular_decomp} and \ref{fig:angular_decomp_3D}, respectively.
	
	Since the overall amplitude of $\tilde{C}(\omega,\theta)$  is suppressed at low frequency, the amplitude  $\tilde{c}_{\ell m}(\omega)$ of different $(\ell,m)$ modes is also suppressed as shown in Fig.~\ref{fig:angular_decomp_3D}. In Fig.~\ref{fig:angular_decomp}, one can also observe that the amplitude $c_{\ell m}$ falls off more slowly with $\ell$ in the case with an IR cutoff since low $\ell$ modes are more sensitive to this IR cutoff and hence are more suppressed. As noted previously, our model with the IR cutoff better agrees with the results in \cite{Verlinde_Zurek_2019_1}, though one should remain cautious until our model has been fully mapped to a UV-complete theory.
	
	\subsection{Existing constraints and future projections}
	\label{sec:experimental_constraints}
	
	In an effort to detect high frequency gravitational waves and quantum gravity signatures, several laboratory-sized interferometer experiments have been implemented to accurately detect tiny spacetime perturbations. The constraints from these experiments are often reported as upper limits on the one-sided noise strain $\sqrt{S_{h}(f)}$ of the photon round-trip time, obtained by analysing interference patterns. For stationary signals, the strain is defined as~\cite{Chou_2017, Moore_2014}
	\begin{equation}
       \sqrt{S_h^{(n)}(f)} = \sqrt{2\int_{-\infty}^{\infty}\left\langle\frac{\Delta L(\tau)}{L}\frac{\Delta L(0)}{L}\right\rangle e^{-2\pi if\tau}d\tau} \, ,
	\end{equation}
	which has units of Hz$^{-1/2}$. This is related to Eq.~\eqref{eq:C_psd} by Eq.~\eqref{eq:C_T_psd}, {\em i.e.},
	\begin{equation}\label{eq:hC}
        \sqrt{S_{h}(f)} = \sqrt{2\tilde{C}_{\mathcal{T}}\left(\omega,\theta=\frac{\pi}{2}\right)} \, ,
	\end{equation}
	where $\omega=2\pi f$ and we assume a perpendicular arm configuration.
    Our power spectrum in Eq.~\eqref{eq:C_psd} can be parameterized more conventionally by defining
	\begin{equation}\label{eq:alpha_a}
        \alpha \equiv \frac{2\pi}{c_s^2}a \, ,
	\end{equation}
    leading to the peak strain $\sqrt{S_{h}(f_{\mathrm{peak}})}\approx\sqrt{2\alpha l_p}/(4\pi)=\sqrt{\alpha}(2.62\times 10^{-23})$ Hz$^{-1/2}$~\footnote{This is related to the one-sided displacement spectrum by $S_{L}(f)=2L^2\tilde{C}(f)$, which is peaked at $S_{L}(f_{\mathrm{peak}})=\alpha l_pL^2/(8\pi^2)$.}.  Here $\alpha \sim 1$ gives the amplitude of the effect computed in \cite{Verlinde_Zurek_2019_1,Verlinde_Zurek_2019_2}, and should be considered the natural benchmark~\footnote{Since $\alpha=1$ corresponds to $a=c_s^2/(2\pi )$, the finite propagation speed $c_s$ has led us to make a corrected prescription of $\beta=l_p/a= 2\pi l_p/c_s^2$ in Eqs.~\eqref{eq:dos_approx} and \eqref{eq:dos}.   }. 
	
\begin{figure*}[t]
		\centering
		\includegraphics[width=1\linewidth]{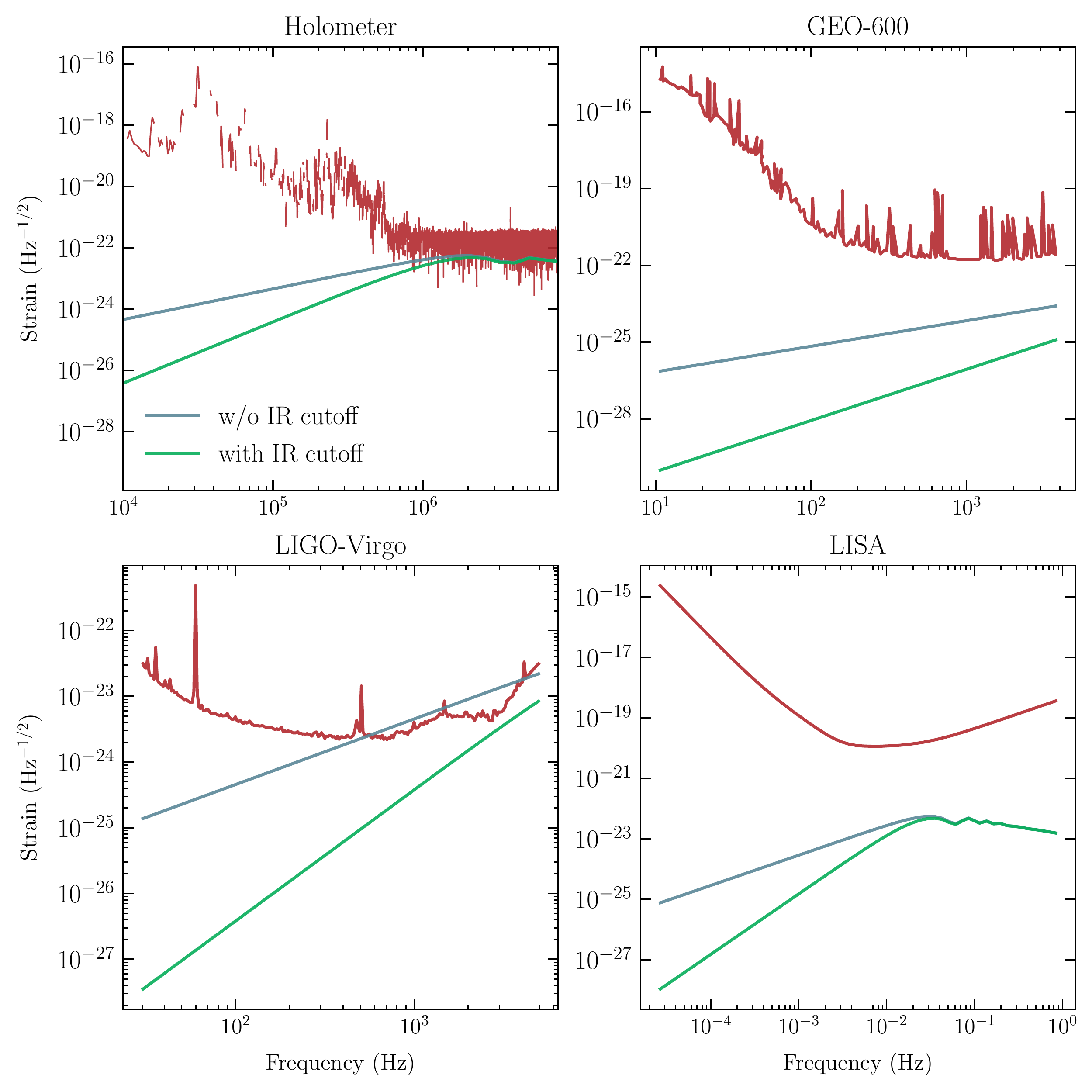}
		\caption{Strain comparison between model predictions (blue and green) and experimental / projection constraints (red). The model curves are computed using Eqs.~\eqref{eq:C_psd},~\eqref{eq:C_T_psd},~\eqref{eq:C_psd_cutoff},~\eqref{eq:hC} and~\eqref{eq:alpha_a} assuming $\alpha=1$, while the experimental curves are extracted from Refs.~\cite{Chou_2017, geo_600, Lee_LIGO_2021, LISA_2021}. The LIGO data shown here are obtained by the Livingston detector, but we note that the Hanford detector yields similar constraints.}
		\label{fig:PSD_comparison} 
	\end{figure*}

	We now compare our predicted strain to the experimental constraints from Holometer~\cite{Chou_2017}, GEO-600~\cite{geo_600}, LIGO~\cite{Lee_LIGO_2021}, and the projected sensitivity from LISA~\cite{LISA_2021}. Since the four interferometers have different arm lengths, the predicted strain from our models will also differ between these experiments. The result assuming $\alpha=1$ with or without the IR cutoff using Eqs.~\eqref{eq:C_psd},~\eqref{eq:C_T_psd},~\eqref{eq:C_psd_cutoff},~\eqref{eq:hC}, and~\eqref{eq:alpha_a} is plotted in Fig.~\ref{fig:PSD_comparison}. As expected, the tightest experimental limit comes from LIGO and Holometer measurements, which at $3\sigma$ significance, are roughly $\alpha\lesssim 3$ and $\alpha\lesssim 0.7$ (with IR cutoff), and $\alpha\lesssim 0.1$ and $\alpha\lesssim 0.6$ (w/o IR cutoff), respectively. On the other hand, our model is out of reach for GEO-600 and LISA.
	
	Caltech and Fermilab are commissioning a joint theoretical and experimental initiative called Gravity from Quantum Entanglement of Space-Time (GQuEST), dedicated to probing the VZ effect proposed in Ref.~\cite{Verlinde_Zurek_2019_1}. This includes the construction of a tabletop optical Michelson interferometer with arm-length $L = 5$ m, with a novel read-out scheme with single photons rather than the usual interference effect.  The advantage of this scheme is that sensitivity beats the standard quantum limit, with signal-to-noise ratio increasing linearly with integration time, rather than the usual square-root dependence. The experiment is projected to be able to constrain $\alpha\lesssim 1$ after 1000 s of background-free integration time, corresponding to a dark count rate of $10^{-3}$ Hz. We expect the constraint on $\alpha$ to tighten {\em linearly} with lower dark count rate and longer integration time.
	
	Some previous works on quantifying spacetime fluctuations (motivated by theories other than the VZ effect) argued that the predicted strain should not be directly compared against experimental constraints such as GEO-600 and LIGO~\cite{Kwon_2016}, since transitional interferometer experiments often utilize Fabry-Perot cavities ({\em e.g.}, LIGO uses Fabry-Perot cavities within each arm, where the average light storage equals to 35.6 light round trips~\cite{LIGO_2009}) to boost the signal-to-noise ratio from astrophysical gravitational waves, while it is unclear whether quantum gravity signals, which are fundamental to spacetime itself, will benefit from additional light-crossings. Here we show that spacetime fluctuations based on Eq.~\eqref{eq:metric_pix} do accumulate over a Fabry-Perot cavity, thus justifying our direct strain comparison with gravitational experiments. A Fabry-Perot Michelson interferometer can be viewed as a linear device that measures the differential single-round-trip phase, $\Delta\Phi=\Phi_1-\Phi_2$ between the two arms --- regardless of whether this phase arises from gravitational waves, displacement of mirrors, or space-time fluctuations. This $\Delta\Phi$ is linearly transferred to the output field $z$, with noise $N$ added:
    \begin{equation}
        z(f)=\mathcal{M}(f)\Delta\Phi(f)+N(f)\,.
    \end{equation}
    In particular, $\mathcal{M}(f)$ contains the build-up (or suppression) of signal due to the Fabry-Perot cavity. 

    We now convert the strain-referred noise spectrum $S_h$ published by LIGO to a spectrum for $\mathcal{T}$.  In obtaining $S_h$ (below 5\,kHz, as shown in Fig.~\ref{fig:PSD_comparison}), LIGO used a long-wave-length approximation, and assumed that the wave has a $+$ polarization (stretching along the $x$ and squeezing along the $y$ direction), and propagating along $z$ --- perpendicular to the detector plane ({\em e.g.}, adopted by Chapter 27.6 of \cite{thorne2017modern}).  In this case, in the Local Lorentz frame of the beam splitter, the first and second mirrors are going to be displaced by $\pm Lh/2$, leading to phase shifts of 
    \begin{equation}
        \Phi_{1,2} = \pm \omega_0 Lh/c
    \end{equation}
    and 
    \begin{equation}
        \Delta\Phi = 2\omega_0 Lh/c\,.
    \end{equation}
    In this way, the $\Delta\Phi$-referred spectrum is related to $S_h$ published by LIGO via
    \begin{equation}
        \sqrt{S_{\Delta\Phi}}
        =\frac{2\omega_0 L}{c}\sqrt{S_h}\,.
    \end{equation} 
    We note that at higher frequencies, and/or for interferometers with longer arms, the conversion from $h$ to $\Phi$ becomes less trivial. In our case, we have
    \begin{equation}
        \Delta\Phi (t)
        =\omega_0[\delta T(t,\mathbf{n}_1)-\delta T(t,\mathbf{n}_2)]
        =\omega_0\mathcal{T}(t,\theta)\,.
    \end{equation}
    We therefore have $\sqrt{S_{\Delta\Phi}}=\omega_0\sqrt{S_\mathcal{T}}$ and thus
    \begin{equation}
        \sqrt{S_{\mathcal{T}}}=\frac{2L}{c} \sqrt{S_h}\,.
    \end{equation}
    This allows us to straightforwardly relate our observable defined in Eqs.~\eqref{eq:corr_T_def} and \eqref{eq:C_T_psd} to the quantity $S_h$ constrained by LIGO. In LIGO, $S_h$ is usually reported as a one-sided spectrum, so we need another factor of $2$ when converting the two-sided spectrum $\tilde{C}_{\mathcal{T}}$ in Eq.~\eqref{eq:C_T_psd} to the one-sided spectrum $S_h$, {\em i.e.,}
    \begin{equation}
        \sqrt{S_h} =\sqrt{S_{\mathcal{T}}}\Big/\left(\frac{2L}{c}\right)
        =\sqrt{2\tilde{C}_{\mathcal{T}}\left(\omega,\theta=\frac{\pi}{2}\right)}\,,
    \end{equation}
    which is consistent with the conversion in Eq.~\eqref{eq:hC}.

	\section{Conclusions}
	\label{sec:conclusions}

    In this paper we have investigated the effects on the fluctuations in the time-of-arrival of a photon in an interferometer, due a scalar field coupled to the metric as in Eq.~\eqref{eq:metric_pix} with an occupation number given by Eq.~\eqref{eq:dos}. This simple scalar field is designed to model the behavior of vacuum fluctuations of the modular energy ({\em e.g.}, Ref.~\cite{Verlinde_Zurek_2019_2}) from shockwave geometries \cite{Verlinde_Zurek_3}.  

    We showed that the interferometer observable had a power spectral density quadratically suppressed $\propto \omega^2$ or $\propto \omega^4$, depending on the IR regulator, at low frequency, and an angular correlation between the interferometer arms consistent with that proposed in Ref.~\cite{Verlinde_Zurek_2019_1}, as expected from shockwave geometries.

    In future work, we plan to more explicitly demonstrate the connection between shockwave geometries and interferometer observables, completing the bridge between the model presented here and the UV-complete theory.

	\section{Acknowledgements}

    We thank Temple He, Allic Sivaramakrishnan, and Jordan Wilson-Gerow for discussions and comments on the draft, and Lee McCuller for comments and help with recasting the Holometer bounds.  We are supported by the Heising-Simons Foundation ``Observational Signatures of Quantum Gravity'' collaboration grant 2021-2817. The work of KZ is also supported by a Simons Investigator award and the U.S. Department of Energy, Office of Science, Office of High Energy Physics, under Award No. DE-SC0011632.  The work of YC and DL is also supported by the Simons Foundation (Award Number 568762), the Brinson Foundation and the National Science Foundation (via grants PHY-2011961 and PHY-2011968). 

	\appendix
	\section{Time Delay in General Metric} \label{appendix:time_delay_general}
	
	In this appendix, we derive the time delay of a generic metric in Eq.~\eqref{eq:metric_general}. There are three effects, from the clock rate, the mirror motion, and the light propagation.  Only when summing all three do we obtain the gauge invariant observable.
 
 We start by computing the clock's rate. Since $g_{tt}=-(1-\mathcal{H}_0)$, to the leading order, the proper time differs from the coordinate time by
	\begin{equation}
		\frac{d\tau}{dt}\approx1-\frac{1}{2}\mathcal{H}_0\,.
	\end{equation}
	Thus, for a clock with radial position $r$ when there is no metric fluctuation, the difference $\delta\tau$ between the proper time and the coordinate time from $t=t_1$ to $t=t_2$ is
	\begin{equation}
		\delta\tau(t_1,t_2,r)
		=-\frac{1}{2}\int_{t_1}^{t_2}dt'\;\mathcal{H}_0(t',r)\,.
	\end{equation}
	
	To account for the mirror's motion, we consider the geodesic equation of the mirror
	\begin{equation}
		0=\frac{d^2x^{\mu}}{d\tau^2}+\Gamma^{\mu}_{\alpha\beta}
		\frac{dx^{\alpha}}{d\tau}\frac{dx^{\beta}}{d\tau}
		\approx\frac{d^2x^{\mu}}{d\tau^2}+\Gamma^{\mu}_{tt}+
		\Gamma^{\mu}_{ti}v^{i}+\cdots\,.
	\end{equation} 
	Since the velocity of the mirror $v^{i}\ll1$, to the leading order, $\frac{d^2r}{dt^2}\approx-\Gamma^{r}_{tt}$. Using $\Gamma_{\alpha\beta}^{\mu}
	=\frac{1}{2}\eta^{\mu\nu}(\partial_{\alpha}h_{\beta\nu}
	+\partial_{\beta}h_{\alpha\nu}-\partial_{\nu}h_{\alpha\beta})$, we get
	\begin{equation} 
		\Gamma^{r}_{tt}
		=\partial_th_{tr}-\frac{1}{2}\partial_rh_{tt}
		=\partial_t\mathcal{H}_1-\frac{1}{2}\partial_r\mathcal{H}_0\,,
	\end{equation}
	so for a mirror at radius $r$ when there is no metric fluctuation, its radial position $r_{\mirror}$ at coordinate time $t$ is
	\begin{equation}
		r_{\mirror}(t,r)\approx\int^{t} dt'\;\int^{t'} dt''\;
		\left[\frac{1}{2}\partial_r\mathcal{H}_0(t'',r)-\partial_{t''}\mathcal{H}_1(t'',r)\right]\,.
	\end{equation}
	
	For the light propagation, the geodesic equation of outgoing light is
	\begin{equation}
		\frac{dt^{\out}}{dr}
		\approx1+\frac{1}{2}\left(\mathcal{H}_0+\mathcal{H}_2
		+2\mathcal{H}_1\right)
		\equiv1+\frac{1}{2}\mathcal{H}^{\out}\,,
	\end{equation}
	and for ingoing light,
	\begin{equation}
		\frac{dt^{\In}}{dr}
		\approx-1-\frac{1}{2}\left(\mathcal{H}_0+\mathcal{H}_2
		-2\mathcal{H}_1\right)
		\equiv-1-\frac{1}{2}\mathcal{H}^{\In}\,.
	\end{equation}
	
	In total, the proper time $T^{\out}$ the light beam takes to reach the mirror is
	\begin{equation} \label{eq:time_delay_out}
		\begin{aligned}
			T^{\out}
			\approx& \;\int_{0+r_{\mirror}(0,0)}^{L+r_{\mirror}(L,L)}dr\;
			\left[1+\frac{1}{2}\mathcal{H}^{\out}(r,r)\right]
			+\delta\tau(0,L,0) \\
			\approx& \;L+r_{\mirror}(L,L)-r_{\mirror}(0,0)+\delta\tau(0,L,0) \\
			& \;+\frac{1}{2}\int_{0}^{L}dr\;\mathcal{H}^{\out}(r,r)\,. \\
		\end{aligned}
	\end{equation}
	Similarly, for the ingoing light beam,
	\begin{equation} \label{eq:time_delay_in}
		\begin{aligned}
			T^{\In}
			\approx& \int_{L+r_{\mirror}(L,L)}^{0+r_{\mirror}(2L,0)}dr\;
			\left[-1-\frac{1}{2}\mathcal{H}^{\In}(2L-r,r)\right]
			+\delta\tau(L,2L,0) \\
			\approx& \;L+r_{\mirror}(L,L)-r_{\mirror}(2L,0)+\delta\tau(L,2L,0) \\
			& \;+\frac{1}{2}\int_{0}^{L}dr\;\mathcal{H}^{\In}(2L-r,r)\,.
		\end{aligned}
	\end{equation}
	Then the total time delay $T$ is given by summing up Eqs.~\eqref{eq:time_delay_out} and \eqref{eq:time_delay_in}, $T=T^{\out}+T^{\In}$.

	\section{Gauge Invariance of Time Delay} \label{appendix:gauge_invariance}
	
	In this appendix, we show that the total time delay $T=T^{\out}+T^{\In}$, where $T^{\out}$ and $T^{\In}$ are defined in Eqs.~\eqref{eq:time_delay_out} and \eqref{eq:time_delay_in}, of the light beam traveling a round trip is a gauge invariant quantity. Since the $t-r$ sector of any metric, {\em e.g.}, Eq.~\eqref{eq:metric_general}, will only be affected by the gauge transformations of coordinate $t$ or $r$, we will show that $T$ is invariant under these two types of gauge transformations.
	
	\subsection{Gauge transformations of coordinate t}
	
	First, let's consider gauge transformations $x^{\mu}\rightarrow x^{\mu}+\xi^{\mu}$, where only $\xi_{t}\neq0$, so the metric becomes
	\begin{equation}
		\begin{aligned}
			ds^2
			=& \;-(1-\mathcal{H}_0+2\partial_t\xi_t)dt^2
			+(1+\mathcal{H}_2)dr^2 \\
			& \;+2(\mathcal{H}_1-\partial_r\xi_t)dtdr+\cdots\,.
		\end{aligned}
	\end{equation}
	Since $h_{tt}$ is modified, $\frac{d\tau}{dt}\rightarrow\frac{d\tau}{dt}
	+\frac{1}{2}\partial_t\xi_t$, the difference between the proper time and the coordinate time becomes
	\begin{equation}
		\delta\tau(t_1,t_2,r)
		\rightarrow\delta\tau(t_1,t_2,r)+\xi_t(t_2,r)-\xi_t(t_1,r)\,.
	\end{equation}
	The geodesics equations of light beam are modified into
	\begin{align}
		& \frac{dt^{\out}}{dr}\approx
		1+\frac{1}{2}\left(\mathcal{H}^{\out}-2\partial_t\xi_t
		-2\partial_r\xi_t\right)\,, \\
		& \frac{dt^{\In}}{dr}\approx
		-1-\frac{1}{2}\left(\mathcal{H}^{\In}-2\partial_t\xi_t
		+2\partial_r\xi_t\right)\,.
	\end{align}
	For mirror's motion, let's define
	\begin{align}
		& \delta r^{\out}\equiv r_{\mirror}(L,L)-r_{\mirror}(0,0)\,,\\
		& \delta r^{\In}\equiv r_{\mirror}(L,L)-r_{\mirror}(2L,0)\,.
	\end{align} 
	Since $\Gamma^{r}_{tt}\rightarrow\Gamma^{r}_{tt}
	-\partial_t\partial_r\xi_t+\partial_r\partial_t\xi_t=\Gamma^{r}_{tt}$ remains unchanged, $\delta r^{\out}_{\mirror}
	\rightarrow\delta r^{\out}_{\mirror}$ and $\delta r^{\In}_{\mirror}
	\rightarrow\delta r^{\In}_{\mirror}$.
	In total,
	\begin{equation}
		\begin{aligned}
			T^{\out}
			\rightarrow& \;T^{\out}+\xi_t(L,0)-\xi_t(0,0)
			-\int_{0}^{L}dr\;\left(\partial_t\xi_t+\partial_r\xi_t\right)\vert_{t=r} \\
			=& \;T^{\out}+\xi_t(L,0)-\xi_t(L,L)\,,
		\end{aligned}
	\end{equation}
	\begin{equation}
		\begin{aligned}
			T^{\In}
			\rightarrow& \; T^{\In}+\xi_t(2L,0)-\xi_t(L,0)
			+\int_{0}^{L}dr\;\left(\partial_r\xi_t-\partial_t\xi_t\right)\vert_{t=2L-r} \\
			=& \;T^{\In}-\xi_t(L,0)+\xi_t(L,L)\,,
		\end{aligned}
	\end{equation}
	so the total time delay of a round trip $T\rightarrow T$
	under the gauge transformation of coordinate $t$.
	
	\subsection{Gauge transformations of coordinate r}
	
	Next, let's consider gauge transformations $x^{\mu}\rightarrow x^{\mu}+\xi^{\mu}$ with $\xi_{r}\neq0$ only. The metric then becomes 
	\begin{equation}
		\begin{aligned}
			ds^2
			=& \;-(1-\mathcal{H}_0)dt^2
			+(1+\mathcal{H}_2-2\partial_r\xi_r)dr^2 \\
			& \;+2(\mathcal{H}_1-\partial_t\xi_r)dtdr+\cdots\,.
		\end{aligned}
	\end{equation}
	The relation between the proper time and the coordinate time remains unchanged. The ingoing and outgoing light's geodesics are modified to be
	\begin{align}
		& \frac{dt^{\out}}{dr}\approx
		1+\frac{1}{2}\left(\mathcal{H}^{\out}-2\partial_r\xi_r
		-2\partial_t\xi_r\right)\,, \\
		& \frac{dt^{\In}}{dr}\approx
		-1-\frac{1}{2}\left(\mathcal{H}^{\In}-2\partial_r\xi_r
		+2\partial_t\xi_r\right)\,.
	\end{align}
	$\Gamma^{r}_{tt}$ now becomes $\Gamma^{r}_{tt}\rightarrow\Gamma^{r}_{tt}
	-\partial^2_t\xi_r$, so
	\begin{align}
		& \delta r^{\out}_{\mirror}
		\rightarrow\delta r^{\out}_{\mirror}+\xi_r(L,L)-\xi_r(0,0)\,, \\
		& \delta r^{\In}_{\mirror}
		\rightarrow\delta r^{\In}_{\mirror}+\xi_r(L,L)-\xi_r(2L,0)\,.
	\end{align}
	Then, in total,
	\begin{equation}
		\begin{aligned}
			T^{\out}
			\rightarrow& \;T^{\out}+\xi_r(L,L)-\xi_r(0,0)
			-\int_{0}^{L}dr\;\left(\partial_r\xi_r+\partial_t\xi_r\right)\vert_{t=r} \\
			=& \;T^{\out}\,,
		\end{aligned}	
	\end{equation}
	\begin{equation}
		\begin{aligned}
			T^{\In}
			\rightarrow& \;T^{\In}+\xi_r(L,L)-\xi_r(2L,0)
			-\int_{0}^{L}dr\;\left(\partial_r\xi_r-\partial_t\xi_r\right)\vert_{t=2L-r} \\
			=& \;T^{\In}\,,
		\end{aligned}	
	\end{equation}
	so $T$ also remains invariant under the gauge transformation of coordinate $r$. Thus, we have shown that $T$ is a gauge invariant quantity.

	\bibliographystyle{apsrev4-1}
	\bibliography{reference}
	
\end{document}